\documentstyle[11pt]{article}\textheight 230mm\textwidth 150mm
            \newcommand{\be}{\begin{eqnarray}}
            \newcommand{\ee}{\end{eqnarray}}
           \newcommand{\eel}[1]{\label{#1}\end{eqnarray}}
\newcommand{\e}[1]{\label{e:#1}\end{eqnarray}}
     \newcommand{\eg}{{\em e.g.\ }}
            \newcommand{\ie}{{\em i.e.\ }}
            \newcommand{\ga}{{\gamma}}
 
            \newcommand{\la}{{\lambda}}
            
            \newcommand{\del}{{\delta}}

           \newcommand{\ra}{{\rightarrow}}
 
            \newcommand{\lra}{{\leftrightarrow}}
            
            \newcommand{\pet}{{\cal P}}
\newcommand{\bapet}{\bar{\cal P}}
 \newcommand{\tal}{\tilde{\alpha}}
 \newcommand{\tbeta}{\tilde{\beta}}
 \newcommand{\tga}{\tilde{\gamma}}
 \newcommand{\tdel}{\tilde{\delta}}
 \newcommand{\tpet}{\tilde{\cal P}}
 \newcommand{\tca}{\tilde{{\cal C}}}
 \newcommand{\tpe}{\tilde{p}}

\newcommand{\ca}{{\cal C}}
\newcommand{\baca}{\bar{\cal C}}
            \newcommand{\tpsi}{\tilde{\psi}}
            \newcommand{\beq}{\begin{quote}}
            \newcommand{\eq}{\end{quote}}
            
            \newcommand{\al}{\alpha}
            \newcommand{\ben}{\begin{enumerate}}
            \newcommand{\een}{\end{enumerate}}
            \newcommand{\bit}{\begin{itemize}}
            \newcommand{\ei}{\end{itemize}}
    	\newcommand{\nn}{\nonumber}
            \newcommand{\r}[1]{(\ref{e:#1})}
            \newcommand{\edfl}[1]{\Label{#1}\end{df}}

\newcommand{\vb}{{\cal h}}
\newcommand{\hb}{{\cal i}}

\newcommand{\cA}{{\cal A}}

\newcommand{\dagg}{^{\dag}}
\newcommand{\sign}{\mbox{sign}}
\newcommand{\bett}{{\bf 1}}
\newcommand{\dif}{{\partial}}
\newcommand{\halv}{\frac{1}{2}}
\newcommand{\half}{\frac{1}{2}}

\begin{document}
\begin{titlepage}
\noindent
G\"{o}teborg ITP 98-16\\
October 16, 1998\\
(Revised  August 26, 1999)

\vspace*{5 mm}
\vspace*{35mm}
\begin{center}{\LARGE\bf Basics of
BRST quantization\\ on inner product
spaces}\end{center} \vspace*{3 mm} \begin{center} \vspace*{3 mm}

\begin{center}Robert
Marnelius\footnote{E-mail: tferm@fy.chalmers.se} and
 Niclas Sandstr\"om\footnote{E-mail: tfens@fy.chalmers.se}
 \\ \vspace*{7 mm} {\sl
Institute of Theoretical Physics\\ Chalmers University of Technology\\
G\"{o}teborg University\\
S-412 96  G\"{o}teborg, Sweden}\end{center}
\vspace*{25 mm}
\begin{abstract}
There is an elaborated abstract form of BRST quantization on
inner product spaces within the
operator formalism which leads to BRST invariant
states of the form $$|ph\hb=e^{[Q,
\psi]} |\phi\hb$$ where  $\psi$ is a
gauge fixing fermion, and where $|\phi\hb$ is
a  BRST invariant state  determined by simple hermitian conditions.
These state representations are
closely related to the path integral formulation.
Here we analyse the basics of this approach  in  detail.
The freedom in the choice
of
$\psi$ and $|\phi\hb$ as well as their properties
under gauge transformations are explicitly determined for
simple abelian models. In all considered cases
SL(2,R) is shown both to be a natural extended gauge
symmetry and to be useful to
determine
$|ph\hb$. The results are also applied to nonabelian
models.\end{abstract}\end{center}\end{titlepage}

\setcounter{page}{1}
\setcounter{equation}{0}
\section{Introduction and results.}
The operator
version and the path integral version of BRST quantization look quite different.
The reason is that the standard operator version does not take into account
all consequences of  the necessity
to work on inner product spaces. Therefore, one of us has been  developing
a BRST
quantization within the operator formalism in which general conditions are
extracted for BRST invariant states to be inner product states
\cite{Simple}-\cite{Time}. Only when these conditions are satisfied do we have a
precise connection with the path integral representation. In fact, this more
elaborated operator version  provides for a precise interpretation of
BRST quantization within the path integral approach which is not exactly
the naive
correspondence \cite{Path}. In the present paper we give a coherent picture of
this  operator version of BRST quantization on inner product spaces.
All basic  properties are explained in detail and explicitly demonstrated
to be true
for simple abelian gauge models. Previous analyses are thereby considerably
extended.
We investigate more general gauge fixings in simple abelian models from a global
point of view, as well as properties under global gauge transformations in these
models. We determine the
exact conditions under which a gauge fixing is satisfactory as well as its
global
uniqueness. These are basic properties underlying any
gauge theory since there always
exists a local abelianization
\cite{Abel}. The only additional feature in a more general gauge theory is its
topological properties which are specific for each model.  All operator formulas
obtained are abstract and formal. To find a precise result and a precise
corresponding path integral representation we have to choose a consistent state
space representation and this we do for the considered models here. The general
formulas only allow for specific classes of representations
which are  referred to as
quantization rules of the basic variables. These rules
automatically lead to precise,
satisfactory path integral representations, which are expected to include all
known rigorous results of BRST quantization within its the path integral
version. This has been shown to be the case for all treated models
including the ones
in this paper.  However, in some
cases new representations have also been found. In fact, a given model  can have
inequivalent  representations  (see
\eg
\cite{Proper}) which means that it may be some physics also in the choice
of state
space representations.

The basic ingredients in the operator version of BRST quantization of
general gauge
theories are the odd, hermitian,  and nilpotent BRST charge operator $Q$ and the
even, antihermitian ghost number operator $N$ \cite{KO}. They satisfy
\be
&&Q^2=0, \quad [N, Q]=Q.
\e{00}
Ghost numbers of operators $A$ and states $|\phi\hb$, denoted $n_A$ and $n$, are
defined by $[A, N]=n_AA$ and $N|\phi\hb=n|\phi\hb$. Hence, $Q$ has ghost
number plus
one according to \r{00}. Usually one also has an even, hermitian Hamiltonian
operator $H$ with ghost number zero which commutes with $Q$. However, any gauge
theory with a nontrivial Hamiltonian may be embedded in a corresponding
reparametrization invariant formulation with a vanishing Hamiltonian. We discard
therefore Hamiltonians in the following. Some details how
they may be introduced into
the formalism are given in \cite{Time}.

The basic condition in BRST quantization is that the
physical degrees of freedom are
described by BRST invariant states, \ie by states satisfying
\be
&&Q|ph\hb=0
\e{01}
where $|ph\hb$ is required to be decomposable into eigenstates of the ghost
number
operator $N$. Formally, there are solutions of \r{01}
of the form $Q|\chi\hb$ for any
state $|\chi\hb$. Since $Q|\chi\hb$ formally are
zero norm states, the true physical
state space is $Ker Q/Im Q$. In order for the last statements to be true it is
absolutely necessary that all states belong to an inner product space in
which all
inner products are well defined. This is of course
well known. However, what is not so
well known is that the last conditions allow us to
derive more precise formulas which
have general solutions which may be specified  precisely. This is described
below.

The first approach to  BRST quantization on inner product spaces was based
on the
condition that the hermitian BRST charge $Q$ must be possible to decompose as
follows
 \be
&&Q=\del+\del^{\dag},\;\;\;\del^2=0,\;\;\;[\del, \del^{\dag}]_+=0,
\e{1}
and that the physical inner product states are solutions to the conditions
 \be
&&\del|ph\hb=\del^{\dag}|ph\hb=0.
\e{2}
(See \cite{Princ,Bigrad,Simple}.)
These conditions severely restrict the way
the model is represented in $Q$ as well as
the solutions of \r{01}. In fact, within the BFV prescription for
$Q$ given in
\cite{BFV,BF1} the condition
\r{1} necessarily requires an even number of first class constraints.
This is also what one has in
the standard BFV-formulation with dynamical
Lagrange multipliers and antighosts (the
so called non-minimal sector). A typical example is Yang-Mills theories. Within
standard  BFV-formulation
it has so far been shown that there indeed exists a $\del$-operator
satisfying the
properties \r{1} for Lie group theories
\cite{Simple,Gauge}.   This
$\del$-operator  was  constructed explicitly
in \cite{Simple,Gauge} which also made it
possible to solve the conditions \r{2}. The solutions turned out to have the
form
\be
&&|ph\hb=e^{[Q, \psi]}|\phi\hb,
\e{3}
where $\psi$ is a hermitian fermionic gauge fixing operator with ghost
number minus
one, and where $|\phi\hb$ is a simple BRST invariant state determined by a
set of
hermitian operators and is decomposable into eigenstates of the ghost number
operator. (The conditions determining
$|\phi\hb$ appear as boundary conditions in the
corresponding path integral representation \cite{Path}.)
$|\phi\hb$ is not a well defined inner
product state by itself. The inner product is
only well defined with the hermitian regulator or gauge fixing factor
$e^{[Q,
\psi]}$ present. ($e^{[Q,
\psi]}$ has ghost number zero.) Locally $\vb ph|ph\hb$ is
independent of $\psi$ and this should also be true globally for
topologically trivial models. However, even for topologically trivial models there
are two distinct sectors characterized by $\psi$ and $-\psi$. Sometimes
they have
opposite norms as will be shown in section 5.  The formula
\r{3} has been applied to several models
\cite{Proper,SL2R}, and
it has been shown to provide for a natural connection between operator
quantization
and the path integral formulation  \cite{Path}. The decomposition \r{1} is not
unique. If \r{1} is true then we also have
 \be
&&Q=\del'+\del^{\prime\dag},\;\;\;\del^{\prime 2}=0,\;\;\;[\del',
\del^{\prime\dag}]_+=0,\nn\\ &&\del'\equiv U\del U\dagg,\quad [Q, U]=0,
\e{02}
where $U$ is a unitary operator. In fact, from \r{2} we find  that
 \be
&&\del'|ph'\hb=\del^{\prime\dag}|ph'\hb=0\quad\Rightarrow\quad |ph'\hb=U|ph\hb,
\e{03}
which expresses the fact that the BRST invariant states are determined up
to unitary
BRST invariant
 transformations. From \r{3} we find then
\be
&&|ph'\hb=e^{[Q, \psi']}|\phi'\hb,\quad \psi'\equiv U\psi U\dagg,\quad
|\phi'\hb\equiv U|\phi\hb.
\e{04}
If $U$ is of the form $U\equiv e^{i[Q, \tpsi]}$,
where $\tpsi$ is a hermitian, fermionic operator with ghost number minus
one exactly
like $\psi$, then  $U$ represents a unitary gauge
transformation. This gauge group is much larger than the original one but
has the
original one as a subgroup.
 All possible $|\phi\hb$-states
and all possible gauge fixing fermions $\psi$ which can be reached by means
of such
$U$-operators defines  a gauge equivalent sector. Here we investigate this gauge
equivalence in detail and show that only the
original gauge group acts effectively on
$|\phi\hb$.

The second approach to BRST quantization on inner product states is based on the
results of \cite{Solv}. All solutions of the
BRST condition \r{01} are here assumed to
be of the form \r{3} where the $|\phi\hb$-states
are determined by conditions of the
form
\be
&&B_i|\phi\hb=0,\quad B_i\equiv[Q, C_i],
\e{05}
where $\{B_i\}$ is the maximal set of independent
simple hermitian operators of this
form. (The appropriate number of $B_i$ operators is one-fourth of the
unphysical degrees of freedom in phase space.) The set
$\{B_i\}$ must be maximal in order to imply
$Q|\phi\hb=0$ which then implies
\r{01}. The number of
$B_i$-operators must be even. In fact, half of the
$B_i$-operators should be bosonic and half fermionic ones.  This condition
corresponds to \r{1} (see \r{07} below). This condition is also natural from the
properties of a Dirac quantization since a bosonic
condition of the form \r{05} leads
to an infinite norm (if the spectrum of
$B_i$ is continuous) while a fermionic condition of the form
\r{05} leads to a zero norm. Hence, the norm of the $|\phi\hb$-state in
\r{3} is a
product of equally many zeros as infinities which
may be reduced to a definite finite
value by means of the regulator or gauge fixing factor $e^{[Q, \psi]}$.
Notice that
\r{05} requires
\be
&&[B_i, B_j]=C_{ij}^{\;\;k}B_k
\e{06}
 for consistency. ($C_{ij}^{\;\;k}$ are BRST invariant and commute with
$B_l$.) This
in turn requires the $B_i$-operators and the corresponding $C_i$-operators
in \r{05}
to be in involution. In section 4 we show that for simple abelian gauge
theories in
arbitrary linear gauges the two approaches are equivalent. The connection
between
\r{05} and
\r{2} follows according to
\r{602} in section 4 from relations like
\be
&&\del =A\Big(e^{[Q, \psi]}B_ae^{-[Q, \psi]}\Big)\dagg\Big(e^{[Q,
\psi]}B_a'e^{-[Q,
\psi]}\Big),
\e{07}
where $B_a$ and $B'_a$ are the bosonic and fermionic parts of the
$B_i$-operators
($B_i=\{B_a, B'_a\}$). $A$ is a bosonic factor which commutes with the
other factor
in $\del$. The conditions \r{2} yield then always \r{3} where $|\phi\hb$
satisfies
\r{05}. However, it is doubtful that the expression \r{07} satisfies the
properties
\r{1} for all possible $B$-operators and all possible gauge fixings $\psi$. The
general relation between \r{1},\r{2} and \r{3},\r{05} remains to be
investigated.

The BRST invariant states in \r{3},\r{03}, and \r{04} contain zero norm
states of the
form
$Q|\chi\hb$. A simple way to determine or project out the BRST singlets was given in
\cite{Solv}. The BRST singlets $|s\hb$ are BRST invariant inner product states
representing the true physical degrees of freedom of the BRST cohomology
($|s\hb\in
\mbox{Ker} Q/\mbox{Im} Q$,  $|ph\hb\in\mbox{Ker} Q$).
The basic condition used in \cite{Solv} was that the BRST singlets, $|s\hb$, are
inner product states provided they are determined by nonhermitian BRST doublet
operators which are in involution in such a way that they together with the
corresponding hermitian conjugate BRST
doublets form BRST quartets in a general
sense (cf \cite{KO}).  In \cite{Solv}
 it was shown that the BRST singlets have
the simple representation
\be
&&|s\hb=e^{[Q, \psi]}|\phi\hb_s
\e{4}
for general gauge theories with finite number of
degrees of freedom. (The generalization to infinite degrees of freedom should be
straight-forward.) $\psi$ is the same gauge fixing fermion as in
\r{3}. However,
$|\phi\hb_s$ is  determined by the conditions
\be
&&D_r|\phi\hb_s=0,
\e{5}
where $D_r$ is a maximal set of {hermitian}
BRST  doublet operators
 in involution.    Since $D_r\equiv\{B_i\equiv[Q, C_i], C_i\}$ and since
there are
equally many bosonic as fermionic $B_i$-operators, there are equally many
bosonic as
fermionic operators in the BRST doublets $D_r$ and the total number of BRST
doublets is a multiple of four.
The conditions
\be
&&C_i|\phi\hb_s=0
\e{08}
in \r{5} are gauge fixing conditions to \r{05} (cf \cite{Aux}). Like for the
$B_i$-operators there are equally many bosonic as
fermionic $C_i$-operators. Eq.\r{08}
not only implies gauge fixing in the usual sense but also ghost fixing. In fact,
$|\phi\hb_s$ has ghost number zero.
 The precise basic
criterion for
$|s\hb$-states to be
BRST singlets  is that $D'_r$ defined by
\be
&&D'_r\equiv e^{[Q, \psi]}D_re^{-[Q, \psi]}
\e{6}
is such that $D'_r$ and $(D'_r)\dagg$ form BRST quartets, \ie
\be
&&[D'_r, (D'_s)\dagg]\;\; \mbox{\small is an invertible matrix operator}.
\e{7}
Note that the regulator factor $e^{[Q, \psi]}$ is hermitian.
Since \r{4}
has been shown to exist for all gauge theories which have the general
BFV-form of
$Q$ \cite{Solv}, also
\r{3} where $|\phi\hb$ satisfies \r{05} should generalize to all kinds of gauge
theories.

In the above  formulations of BRST quantization on
inner product spaces the
physical states are determined purely algebraically. They are
 therefore formal since
it remains to find an explicit realization of the basic
operators and states consistent
with  these results. The approach is therefore different from
 the way mathematicians
attack cohomology problems. They prescribe the basic operators
and states from the
very beginning. The above approaches leave some freedom for physical
 intuitions to act
in the final explicit realization. Here we make a general investigation of state
representations of the unphysical variables in which the basic variables
even may
have complex eigenvalues.

Some allowed forms of the gauge fixing fermion $\psi$ in \r{3} and
\r{4} were given in
\cite{Solv} and further discussed  in \cite{Geza}. However, in the present
paper we
consider a much more general set of gauge fixing fermions $\psi$. We
investigate in
detail the freedom in the choices of $\psi$ and
$|\phi\hb$ in the formula \r{4} , as well as their
properties under global extended gauge transformations for simple abelian
models. We
discuss gauge equivalences and the nilpotence of $\psi$. (If $\psi$ is
nilpotent
it may be interpreted  as a coBRST charge (see
\cite{Geza})).  The allowed forms of $\psi$
turns out to be in agreement with the allowed forms obtained in
\cite{BFV} within the path integral formulation,
which again demonstrates the close
connection between the two formulations. Some limiting cases are discussed
in detail.

In section 2 the simple abelian models
are presented and the results of
\cite{Geza} are generalized. In section 3
we consider a corresponding general class of gauge
transformations and their properties.
In section 4 we give the existence conditions
for BRST invariant inner product
spaces and compare the two approaches presented above. In
section 5 we define wave function
representations of the BRST singlets and
demonstrate that they are consistent
if half the fundamental hermitian operators are
chosen to have imaginary eigenvalues. This is investigated in its most
general form.
In section 6 we consider the still more general
gauge fixing fermions $\psi$ which
are allowed for bosonic and fermionic gauge theories and the
properties of the corresponding BRST
singlets. We derive the conditions under which these $\psi$'s
may be derived from the
$\psi$'s in section 2 by means of unitary gauge transformations.
The main structure of
section 2 is shown to be  retained. In section 7 we show that our  formulas
also may
be applied to nontrivial models. Finally the paper is concluded
 in section 8. In two appendices we display the
basic formulas used in the text and indicate their derivations.

\setcounter{equation}{0}
\section{Physical states in simple abelian models}
Consider as in \cite{Geza}  simple abelian models whose
hermitian BRST charge operator is given by
\be
&&Q=\ca^ap_a+\bapet^a\pi_a,
\e{21}
where $p_a$ and $\pi_a$ are hermitian conjugate momenta to the
hermitian bosonic coordinates $x^a$ and
$ v^a$ respectively, and $\ca^a$ and $\bapet^a$ are hermitian
fermionic operators conjugate to the hermitian operators
$\pet_a$ and $\baca_a$ respectively. The index $a=1,\ldots,n < \infty$
is assumed to be
raised and lowered by a real, symmetric metric $g_{ab}$. (Up to section 6 the
explicit form of $g_{ab}$
 never enters  our formulas
 since our treatment is manifestly covariant. In section 6  $g_{ab}$
is  restricted to be real, symmetric and
constant.) The fundamental nonzero commutators are
\be &&[x^a, p_a]_-=i\del^a_b,\;\;\;[ v^a,
\pi_b]_-=i\del^a_b,\;\;\;[\ca^a, \pet_b]_+=\del^a_b,\;\;\;[\baca^a,
\bapet_b]_+=\del^a_b.
\e{22}
One may think of \r{21} as the BRST charge operator of an abelian
bosonic gauge theory where $p_a$
are the gauge generators, $ v^a$ the Lagrange multipliers, and $\ca^a$ and
$\baca_a$ the ghosts and
antighosts respectively. Alternatively one may view it as the BRST
charge of a fermionic gauge
theory with bosonic ghosts $p_a$ and
antighosts $ v^a$, or a mixture of these two
interpretations.

Following the approach to BRST
quantization on inner product spaces as proposed in
\cite{Solv}, we first look for a maximal set of hermitian BRST doublets in
involution. Here there are two natural sets, namely
\be
&&D_{(1)}\equiv\{x^a, \ca^a, \baca_a, \pi_a\},
\;\;\;D_{(2)}\equiv\{ v^a, \bapet^a,
\pet_a, p_a\}.
\e{26}
They are dual in the sense that they together
form  BRST quartets since the
matrix
operator $[D_{(1)}, D_{(2)}]$ is invertible.
 We are therefore led to two natural
choices for $|\phi\hb$-states. We have $|\phi\hb_1$
and $|\phi\hb_2$ determined by the
conditions
\be
&&\ca^a|\phi\hb_1=\pi_a|\phi\hb_1=0,\nn\\
&&\bapet^a|\phi\hb_2=p_a|\phi\hb_2=0,
\e{25}
and
\be
&&x^a|\phi\hb_1=\baca_a|\phi\hb_1=0,\nn\\
&& v^a|\phi\hb_2=\pet_a|\phi\hb_2=0.
\e{251}
The conditions \r{25} may equivalently be written as
\be
&&[Q,x^a]|\phi\hb_1=[Q,\baca_a]|\phi\hb_1=0,\nn\\
&& [Q,v^a]|\phi\hb_2=[Q,\pet_a]|\phi\hb_2=0,
\e{252}
\ie they are of the form \r{05}, while  \r{251} are gauge fixing conditions
of the
form \r{08}.  Notice also that \r{25} implies
$Q|\phi\hb=0$.
Out of $|\phi\hb_1$
and $|\phi\hb_2$ we may construct BRST singlets
according to the rule \r{4}. We are
then led to the following representations:
\be
&&|s\hb_l=e^{[Q, \psi_l]}|\phi\hb_l,\;\;\;l=1,2,
\e{23}
where $\psi_1$ and $\psi_2$ are gauge fixing fermions.
In \cite{Solv,Geza} it was
shown that the choices
\be
&&\psi_1=\al\pet_a v^a,\;\;\;\psi_2=\beta\baca_a x^a,
\e{24}
are satisfactory provided
 the real constants $\al$ and $\beta$
are different from zero,  since only then do the
states \r{23} satisfy the criterion \r{7}.
We notice that $\psi_1$ and $\psi_2$ are
products of the gauge fixing variables to the conditions in \r{25}.  The inner
products
$_l\vb s|s\hb_l$ are independent of $\al$ and
$\beta$ for positive and negative
values separately. However, they are undefined at
$\al=0$ and
$\beta=0$.
In \cite{Geza} it was  shown that the two
singlets in \r{23} are related as follows
\be
&&e^{\al K_1}|\phi\hb_1=e^{-{1\over\al}K_2}|\phi\hb_2,
\e{27}
where
\be
&&K_1\equiv[Q, \pet_a v^a]= v^a p_a+i\pet_a\bapet^a,\nn\\
&&K_2\equiv[Q, \baca_a x^a]=x^a \pi_a+i\baca_a \ca^a.
\e{28}
One may easily check that the both sides in \r{27}
satisfy the same conditions. Note that
the BRST singlets are  unique up to unitary
transformations.

In \cite{Geza} also the more general gauge fixing fermion
\be
&&\psi_1=\psi_2=\al\pet_a v^a+\beta\baca_a x^a
\e{29}
was investigated. It was found that
\be
|s\hb_1=e^{\al K_1+\beta K_2}|\phi\hb_1=e^{\al' K_1}|\phi\hb_1,\;\;\;
|s\hb_2=e^{\al K_1+\beta K_2}|\phi\hb_2=e^{\beta' K_2}|\phi\hb_2,
\e{210}
where
\be
&&\al'=\al\,\frac{\tan{\sqrt{\al\beta}}}{\sqrt{\al\beta}},
\;\;\;\beta'=\beta\,\frac{\tan{\sqrt{\al\beta}}}{\sqrt{\al\beta}}
\e{211}
for $\al\beta>0$ and
\be
&&\al'=\al\,\frac{\tanh{\sqrt{-\al\beta}}}{\sqrt{-\al\beta}},
\;\;\;\beta'=\beta\,\frac{\tanh{\sqrt{-\al\beta}}}{\sqrt{-\al\beta}}
\e{212}
for $\al\beta<0$. From \r{212} it follows that
 $|s\hb_1$ and $|s\hb_2$
 are well defined inner product
states provided $\al'$
and $\beta'$ are non-zero
and finite.   In \cite{Geza} it was also shown that
there are natural representations of
the operators and states that make the inner
products of $|s\hb_{1,2}$ explicitly finite.

According to the BFV-formulation
\cite{BFV,BF1} the general form of the gauge fixing
fermion is
\be
&&\psi=\pet_a\Lambda^a+\baca_a\chi^a
\e{213}
in the case when the simple abelian model
\r{21} is considered to describe a bosonic
gauge theory.
$\Lambda^a$ and $\chi^a$ are then bosonic
gauge fixing variables to the gauge
generators $\pi_a$ and $p_a$. If on the
other hand \r{21} is considered to describe a
fermionic gauge theory, then the general form of $\psi$ is
\be
&&\psi=x^a\tilde{\chi}_a+v^a\tilde{\Lambda}_a,
\e{214}
where $\tilde{\chi}_a$ and $\tilde{\Lambda}_a$
are fermionic gauge fixing variables to
$\ca^a$ and
$\bapet^a$. The forms \r{213} and \r{214}
are different due to the different choices
of ghost number operators together with
the fact that the forms \r{213} and \r{214}
follow from the requirement that $\psi$
must have ghost number minus one. The latter
condition implies that the ghost number
operator  commutes with the regulator
factor $e^{[Q, \psi]}$, a condition
 which always must be true. The ghost number
operators in cases
\r{213} and
\r{214} are
\be
&&N_b=\ca^a\pet_a-\baca_a\bapet^a,\quad N_f=-ip_ax^a-iv^a\pi_a,
\e{215}
respectively.
Note that $N_{b,f}$ is antihermitian, $N_{b,f}^{\dag}=-N_{b,f}$.

The most general form of $\psi$ consistent with both interpretations above,
\ie in
which both ghost number operators in
\r{215} are conserved (\ie commuting with $[Q,
\psi]$), and where
$\Lambda^a$,
$\chi^a$,
$\tilde{\chi}_a$, and
$\tilde{\Lambda}_a$ are chosen to be linear and covariant is
\be
&&\psi=\al\pet_a v^a+\beta\baca_a
x^a+\ga(\baca_av^a-\pet_ax^a)+\del/2(\baca_av^a+\pet_ax^a),
\e{216}
where $\al$, $\beta$, $\ga$, and $\del$ are real parameters. This $\psi$ is
nilpotent, $\psi^2=0$. We notice  that in order for the corresponding
$\Lambda^a$,
$\chi^a$, and  $\tilde{\chi}_a$,
$\tilde{\Lambda}_a$, to be consistent gauge fixing
variables classically, we must have
$(\del/2)^2\neq\al\beta+\ga^2$ ($\Lambda^a$, $\chi^a$ $\lra$ $v^a$, $x^a$
and
$\tilde{\chi}_a$,
$\tilde{\Lambda}_a$ $\lra$  $\baca_a$, $\pet_a$ are then one-to-one).
Although this
condition will appear at several occasions in the following the inner
products will
always be well defined for
$(\del/2)^2=\al\beta+\ga^2$ like in \r{24}.
The gauge fixing fermion \r{216}  yields now
\be
&&[Q, \psi]=\al K_1+\beta K_2+2\ga K_3+\del K_4,
\e{217}
where $K_1$ and $K_2$ are defined in \r{28} and where
\be
&K_3\equiv &\halv[Q,\baca_av^a-\pet_ax^a]=\halv\left(
v^a\pi_a-p_ax^a-i\pet_a\ca^a-i\bapet^a\baca_a\right) =\nn\\
&&=\halv\left(\pi_a v^a-x^ap_a+i\ca^a\pet_a+i\baca_a\bapet^a\right),\nn\\&
K_4\equiv &\halv[Q,\baca_av^a+\pet_ax^a]=\halv\left(
v^a\pi_a+p_ax^a+i\pet_a\ca^a-i\bapet^a\baca_a\right) =\nn\\
&&=\halv\left(\pi_a v^a+x^ap_a-i\ca^a\pet_a+i\baca_a\bapet^a\right).
\e{218}
The operators $K_i$, $i=1,2,3,4$, satisfy a closed algebra:
The operators $K_1$, $K_2$, and $K_3$ satisfy  the algebra
\be
&&[K_1, K_2]=-2iK_3,\;\;\;[K_1, K_3]=iK_1,\;\;\;[K_2, K_3]=-iK_2,
\e{219}
which is an SL(2,R) algebra. (By means of the identification
$\phi_1=1/2(K_2-K_1),\;\;\phi_2=1/2(K_1+K_2),\;\;\phi_3=K_3$, we arrive at
the standard SL(2,R) algebra
$[\phi_i, \phi_j]=i\varepsilon_{ij}^{\;\;\;k}\phi_k$
with the metric Diag($\eta_{ij})=(-1,+1,+1)$.) This was also shown in
\cite{Geza}.
In addition we have
\be
&&[K_4, K_i]=0\quad i=1,2,3,4.
\e{220}
Hence, we may view $K_i$ as generators of SL(2,R)$\otimes$U(1).
The $K_i$ operators satisfy also the properties
\be
&&K_2|\phi\hb_1=K_1  |\phi\hb_2=K_3    |\phi\hb_1=K_3|\phi\hb_2=0,\nn\\
&&K_4|\phi\hb_1=K_4|\phi\hb_2=0.
\e{221}
These properties of $K_4$ may also be understood from the equality
\be
&&K_4=\half i(N_f-N_b),
\e{2211}
since $K_i$ and the $|\phi\hb$-states have ghost number zero with respect
to both
ghost number operators in \r{215}. The above  properties
of $K_i$ and $K_4$ imply now (see appendix A)
\be
&&|s\hb_1=e^{\al K_1+\beta K_2+2\ga K_3+
\del K_4}|\phi\hb_1=e^{\al K_1+\beta K_2+2\ga
K_3}|\phi\hb_1=e^{\al' K_1}|\phi\hb_1,
\nn\\&& |s\hb_2=e^{\al K_1+\beta K_2+2\ga
K_3+\del K_4}|\phi\hb_2=e^{\al K_1+\beta K_2+2\ga
K_3}|\phi\hb_2=e^{\beta'K_2}|\phi\hb_2,
\e{222}
where the second equalities  trivially
follow from \r{220} and the last two equalities
in \r{221}. The last equalities in
\r{222} follow from the formulas \r{a9}-\r{a10}in appendix A.
The parameters
$\al'$ and $\beta'$ are here in general complex  and given by
\be
&&\al'=\al\,\frac{\tan{\sqrt{\al\beta+\ga^2}}}{\sqrt{\al\beta+\ga^2}+
i\ga\tan{\sqrt{\al\beta+\ga^2}}},\nn\\&&
\beta'=\beta\,
\frac{\tan{\sqrt{\al\beta+\ga^2}}}{\sqrt{\al\beta+\ga^2}-
i\ga\tan{\sqrt{\al\beta+\ga^2}}}
\e{223}
for $\al\beta+\ga^2>0$ and
\be
&&\al'=\al\,\frac{\tanh{\sqrt{-\al\beta-\ga^2}}}
{\sqrt{-\al\beta-\ga^2}+
i\ga\tanh{\sqrt{-\al\beta-\ga^2}}},\nn\\&&
\beta'=\beta\,
\frac{\tanh{\sqrt{-\al\beta-\ga^2}}}{\sqrt{-\al\beta-\ga^2}-
i\ga\tanh{\sqrt{-\al\beta-\ga^2}}}
\e{224}
for $\al\beta+\ga^2<0$. Finally for
$\al\beta+\ga^2=0$ these expressions reduce to
\be
&&\al'={\al\over 1+i\ga},\quad \beta'={\beta\over 1-i\ga}.
\e{225}
From
\r{222} it follows that provided the
real parts of $\al'$ and $\beta'$ are non-zero
and finite,   $|s\hb_1$ and $|s\hb_2$ are well defined inner product
states. (Only the real parts contribute
to the inner products.) These formulas reduce
to
\r{211} and
\r{212} for
$\ga=0$ as they should.

\setcounter{equation}{0}
\section{Unitary gauge transformations}
BRST invariant unitary operators,
$U$, transform our singlet states as follow
\be
&&|s\hb'=U|s\hb=e^{[Q, \psi']}
U|\phi\hb,\quad\psi'=U\psi U^{\dagger},
\e{31}
which shows that changes in
$|\phi\hb$-states and gauge fixings are related.
Gauge transformations  are performed by
 BRST invariant operators of the form
\be
&&U=e^{i[Q,\tilde{\psi}]},
\e{32}
where $\tilde{\psi}$ is an odd,
hermitian operator with ghost number minus one
like the gauge fixing fermion $\psi$.
If we only consider gauge transformations which
commute  with the two ghost number
operators in \r{215} then the general covariant
form of $\tilde{\psi}$ is given in
\r{216}. We are are then led to consider
 gauge transformations obtained from  the unitary operator
\be
&&U(a,b,c,d)=\exp{\left(ia K_1+ib K_2+2ic K_3+idK_4\right)}.
\e{33}
This operator performs $SL(2,R)\otimes U(1)$ transformations.
This might be confusing
since we are considering abelian gauge models. If $x^a$ and $v^a$ (or $\pet_a$
and $\baca_a$) were parameters in $\psi$ we would obtain expected abelian gauge
transformations. However, as will be seen below only an abelian subgroup of
\r{33}
acts effectively on the BRST invariant states.

From
\r{a8} in appendix A we find that
\be
&&e^{\al' K_1+\beta' K_2+2\ga' K_3}=U(a,b,c,d)e^{\al K_1+\beta K_2+2\ga
K_3}U\dagg(a,b,c,d),
\e{331}
where
\be
&&\al'=\al(A+cB)^2-\beta a^2B^2-2\ga a(A+cB)B,\nn\\
&&\beta'=-\al b^2 B^2+\beta(A-cB)^2+2\ga b(A-cB)B,\nn\\
&&\ga'=-\al b(A+cB)B+\beta a(A-cB)B+\ga(1+2abB^2),
\e{332}
where in turn $A$ and $B$ are given in \r{a6}-\r{a7}.
Thus, a general gauge fixing factor with $\del=0$ may
be obtained from specific ones
by means of these unitary gauge transformations. Note that
$\al'\beta'+(\ga')^2=\al\beta+\ga^2$ always.

We determine now what
\r{33} does on the $|\phi\hb$-states.
 On $|\phi\hb_1$ and $|\phi\hb_2$ we find
\be
&&|\phi\hb'_1=U(a,b,c,d)|\phi\hb_1=
U(a,b,c,0)|\phi\hb_1=e^{ia'K_1}|\phi\hb_1,\nn\\
&&|\phi\hb'_2=U(a,b,c,d)|\phi\hb_2=
U(a,b,c,0)|\phi\hb_2=e^{ib'K_2}|\phi\hb_2,
\e{34}
where
\be
&&a'=a\,\frac{\tanh{\sqrt{ab+c^2}}}{\sqrt{ab+c^2}-
c\tanh{\sqrt{ab+c^2}}},\nn\\&&
b'=b\,
\frac{\tanh{\sqrt{ab+c^2}}}{\sqrt{ab+c^2}+
c\tanh{\sqrt{ab+c^2}}}
\e{35}
for $ab+c^2>0$ and
\be
&&a'=a\,\frac{\tan{\sqrt{-ab-c^2}}}{\sqrt{-ab-c^2}-
c\tan{\sqrt{-ab-c^2}}},\nn\\&&
b'=b\,
\frac{\tan{\sqrt{-ab-c^2}}}{\sqrt{-ab-c^2}+
c\tan{\sqrt{-ab-c^2}}}
\e{36}
for $ab+c^2<0$. Finally for $ab+c^2=0$ these expressions reduce to
\be
&&a'={a\over 1-c},\quad b'={b\over 1+c}.
\e{37}
Although
 the reduced parameters $a'$ and $b'$ are
infinite for some values of $a,b$ and $c$ in all these cases, the
transformations
\r{34} are well defined even for these values.  Due to the equality
\be
&&e^{iaK_1}|\phi\hb_1=e^{i{1\over a}K_2}|\phi\hb_2.
\e{38}
valid for any parameter $a$
(both sides satisfy the same equations),
 an infinite value of $a'$ or
$b'$ in \r{34} implies $|\phi\hb'_1=|\phi\hb_2$
and $|\phi\hb'_2=|\phi\hb_1$
respectively. Thus, \r{33} is well defined and reduces to abelian gauge
transformations on the states $|\phi\hb_1$ and $|\phi\hb_2$. Below we show
that the
same result may be obtained by an abelian subgroup of \r{33}.

 Consider  the particular abelian unitary gauge operator
\be
&&U_R(\theta)\equiv e^{i\theta(K_1-K_2)}.
\e{39}
It rotates the basic variables an angle $\theta$.
From the above formulas we find
\be
&&U_R(\theta)|\phi\hb_1=e^{i\theta'K_1}|\phi\hb_1, \quad
U_R(\theta)|\phi\hb_2=e^{-i\theta'K_2}|\phi\hb_2,
\e{310}
where $\theta'=\tan\theta$. Now since
$-\infty<\tan\theta<\infty$ it follows that the
general gauge transformations \r{34} on
$|\phi\hb_1$ and $|\phi\hb_2$ may be
represented by  states of the form
\be
&&|\phi\hb_\theta \equiv U_R(\theta)|\phi\hb_1
\e{311}
for some value of the parameter $\theta$ ($-\half\pi<\theta\leq\half\pi$).
  The state \r{311} is annihilated by the hermitian BRST doublets
\be
&&D_\theta\equiv\left\{\begin{array}{ll}
U_R(\theta)x^a U_R(\theta)\dagg,
&U_R(\theta)\ca^a U_R(\theta)\dagg,\\
 U_R(\theta)\baca_a U_R(\theta)\dagg,&
U_R(\theta)\pi_a U_R(\theta)\dagg\end{array}
\right\}=\nn\\&&=\left\{\begin{array}{ll}
x^a\cos{\theta}+ v^a\sin{\theta},
&\ca^a\cos{\theta}+
\bapet^a\sin{\theta},\\ \baca_a\cos{\theta}-\pet_a
\sin{\theta},& \pi_a\cos{\theta}-
p_a\sin{\theta}\end{array}   \right\}
\e{312}
in accordance with \r{25}. The states \r{311} represent therefore the most
general
class of
$|\phi\hb$-states determined by linear, covariant equations. Note that although
$U_R(\theta)$ on operators naturally is defined for $-\pi<\theta\leq\pi$, the
$|\phi\hb$-states require only $-\half\pi<\theta\leq\half\pi$ since
$|\phi\hb_\theta$
is independent of the sign of $D_{\theta}$.

From \r{311} and \r{312} we have
\be
&&|\phi\hb_2=U_R(\pi/2)|\phi\hb_1,
\e{313}
\ie $|\phi\hb_1$ and $|\phi\hb_2$ are related by
a rotation 90 degrees. This shows that the cases 1 and 2 above are unitarily
equivalent. This is also true for the BRST singlets themselves. Since
\be
&&U_R(\theta)K_1U_R^\dag(\theta)=K_1\cos^2\theta
-K_2\sin^2\theta+K_3\sin2\theta,\nn\\
&&U_R(\theta)K_2U_R^\dag(\theta)=K_2\cos^2\theta
-K_1\sin^2\theta+K_3\sin2\theta,\nn\\
&&U_R(\theta)K_3U_R^\dag(\theta)=K_3,
\e{314}
we find
\be
&&U_R(\pi/2)|s\hb_1=U_R(\pi/2)e^{\al K_1}|\phi\hb_1
=e^{-\al K_2}|\phi\hb_2=|s\hb_2.
\e{315}

Although the other SL(2,R) transformations
do not generate other states than \r{311}
from
$|\phi\hb_1$, they do change $K_i$ in
the regulator factor. We have in particular
\be
&&U_S(\rho)=e^{i\rho K_3},
\e{316}
which scales the basic operators since
\be
&&U_S(\rho)K_1U_S^\dag(\rho)=e^{\rho} K_1,\quad
U_S(\rho)K_2U_S^\dag(\rho)=e^{-\rho}K_2.
\e{317}
This combined with $U_S(\rho)|\phi\hb_1=|\phi\hb_1$ implies
that singlets of the form \r{23} with different parameter values are unitarily
equivalent. Note also that
\be
&&U(a,b,c,d)e^{\al K_1}U(a,b,c,d)\dagg|\phi\hb_1=e^{\al'K_1}|\phi\hb_1
\e{318}
where
\be
&&\al'=\al{(A+cB)^2\over 1+ibB(A+cB)}
\e{319}
from \r{a8} and \r{222}. Thus, the state
\r{318} is up to a simple gauge transformation of the
form $U=e^{ia'K_1}$ equal to a
state $e^{\al''K_1}|\phi\hb_1$ where the {\em real} parameter $\al''$ is
equal to
$(A+cB)^2/(1+b^2B^2(A+cB)^2)$. Note also the peculiar identity transformation
\be
&&U_S(\ln \al^2)U_R(\pi/2)e^{\al K_1}|\phi\hb_1=
e^{-{1\over \al} K_2}|\phi\hb_2=e^{\al
K_1}|\phi\hb_1
\e{320}
due to the equality \r{27}.

A third basic abelian unitary gauge operator is
\be
&&U_H(\la)\equiv U(\la,\la,0)=e^{i\la(K_1+K_2)},
\e{321}
which performs hyperbolic transformations, \ie
\be
&&D_\la\equiv\left\{\begin{array}{ll}U_H(\la)x^a U_H(\la)\dagg,
&U_H(\la)\ca^a U_H(\la)\dagg,\\ U_H(\la)\baca_a U_H(\la)\dagg,&
U_H(\la)\pi_a U_H(\la)\dagg\end{array}
\right\}=\nn\\&&=\left\{\begin{array}{ll}
x^a\cosh{\la}+ v^a\sinh{\la},
&\ca^a\cosh{\la}+
\bapet^a\sinh{\la},\\ \baca_a\cosh{\la}-\pet_a
\sinh{\la},& \pi_a\cosh{\la}-
p_a\sinh{\la}\end{array}   \right\}.
\e{322}
From \r{a8} in Appendix A we have also
\be
&&U_H(2\la)K_1U_H^\dag(2\la)=
\cosh^2\la K_1+\sinh^2\la K_2-\sinh 2\la K_3,\nn\\&&
U_H(2\la)K_2U_H^\dag(2\la)=
\cosh^2\la K_2+\sinh^2\la K_1+\sinh 2\la K_3,\nn\\&&
U_H(2\la)K_3U_H^\dag(2\la)=(1-\sinh^2 \la)K_3+\halv
\sinh 2\la(K_1+K_2).
\e{323}
Note that
\be
&&U_H(\la)|\phi\hb_1=U_R(\theta)|\phi\hb_1,\quad
D_\la U_H(\la)|\phi\hb_1=D_\theta U_R(\theta)|\phi\hb_1=0
\e{324}
for $\theta=\arctan(\tanh\la)$. However,
there is no unitary operator of the form
\r{32} which changes the sign of
$\al$ in $|s\hb_1=e^{\al K_1}|\phi\hb_1$ since the existence of such a unitary
operator would mean that $|\phi\hb_1$ is a well
defined inner product state which it
is not. Thus, singlet states with
opposite signs of $\al$  are not gauge equivalent.
There are two options here: either
we may identify states with opposite signs of $\al$ by hand, or one of them
may be excluded (see next section).

\setcounter{equation}{0}
\section{Existence conditions of the inner product solutions}
The basic abstract criterion for the BRST singlets to be inner product states is
given by \r{7}. The BRST doublets $D'_r$ in \r{6} are  here \eg given by
\be
&&\ca^{'a}=e^{[Q,\psi]}\ca^{a}e^{-[Q,\psi]},\quad
\pi'_a=e^{[Q,\psi]}\pi_ae^{-[Q,\psi]},\nn\\
&&\baca^{'a}=e^{[Q,\psi]}\baca^{a}e^{-[Q,\psi]},\quad
x'_a=e^{[Q,\psi]}x_ae^{-[Q,\psi]},
\e{601}
where
\be
&&e^{[Q, \psi]}=e^{\al K_1+\beta K_2+2\ga K_3+\del K_4}.
\e{602}
$\psi$ is the gauge fixing fermion \r{216}. The operators in \r{601} are
explicitly
given in  \r{a2} appendix A.  We find now that the only non-zero
commutators between
$D'_r$ and $(D'_r)\dagg$ are
\be
&[\pi'_a, (x^{'b})\dagg]=\al F e^{i\del}\del_a^b,\quad
[\ca^{'a}, (\baca'_b)\dagg]=i\al F e^{i\del}\del^a_b,
\e{603}
where $\al$ and $\del$ are the parameters in \r{602} and where
\be
&F\equiv\left\{\begin{array}{cc} {\sin
2\sqrt{\al\beta+\ga^2}\over
\sqrt{\al\beta+\ga^2}},&\al\beta+\ga^2>0,\\
{\sinh
2\sqrt{-\al\beta-\ga^2}\over
\sqrt{-\al\beta-\ga^2}},&\al\beta+\ga^2<0,\\
2,& \al\beta+\ga^2=0.
\end{array}\right.,
\e{604}
also in terms of the parameters in \r{602}. The criterion \r{7} requires the
commutators \r{603} to be non-vanishing. Thus, in order for the BRST singlets
$|s\hb_1$ in \r{222} to be inner product states we must have
\be
&&\al F\neq0\quad\Leftrightarrow\quad\al\neq0\mbox{ and }\sin2
\sqrt{\al\beta+\ga^2}\neq0.
\e{605}

If we define the operators $\phi_a, \rho^a, \xi^a, k_a$ by
\be
&&\phi_a\equiv\pi'_a, \quad \rho^a\equiv \ca'^a,\quad\xi^a\equiv{e^{i\del}\over\al
F}x^{'a},
\quad k_a\equiv i{e^{i\del}\over\al
F}\baca'_a,
\e{606}
then  $\phi_a, \rho^a, \xi^a, k_a$ and their hermitian conjugates
commute among themselves
except for the following two commutators
\be
&&[\phi_a, (\xi^b)\dagg]=\del_a^b, \quad [\rho^a, k_b\dagg]=\del^a_b.
\e{607}
The BRST singlets $|s\hb_{1}$ in \r{222} satisfy then
\be
&&\phi_a|s\hb_{1}=\rho^a|s\hb_{1}=\xi^a|s\hb_{1}=k_a|s\hb_{1}=0,
\e{6071}
which means that the states
$|s\hb_{1}$ represent
 a Fock vacuum with respect to these operators and have  well defined inner
products. The complete Fock space
consists then of half positive and half indefinite
metric states (see \eg section 5 in \cite{Geza}). (The same reasoning is
true for
$|s\hb_{2}$ in \r{224} and the doublet set $D_{(2)}$ in \r{26}.) Another
Fock vacuum
is
\be
&&(\phi_a)\dagg|s\hb_{3}=(\rho^a)\dagg|s\hb_{3}=
(\xi^a)\dagg|s\hb_{3}=(k_a)\dagg|s\hb_{3}=0.
\e{6072}
The solutions are here of the form
\be
&&|s\hb_3=e^{-[Q,\psi]}|\phi\hb_1.
\e{6073}

So far we have only considered the
 approach to BRST quantization on inner
product spaces as presented in \cite{Solv} which is for BRST singlets. As was
mentioned in the introduction  there is also a gauge invariant  approach which
yield similar solutions  that are not gauge fixed.  A crucial issue in this
approach is to find the decomposition \r{1}, \ie
$Q=\del+\del\dagg$. As was also mentioned in the introduction  if one such
solution
is found we have a  whole set of solutions of the form
\be
&&\del'=U\del U\dagg, \quad [Q, U]=0
\e{608}
where $U$ is a unitary operator. For the
simple abelian model \r{21} there
is indeed a decomposition \r{1} which is such that the
solutions of \r{2} will contain the
singlets $|s\hb_1$ in \r{222}. It is
\be
&&\del=\frac{ie^{i\del}}{\al F}(\pi'_a)\dagg\ca^{'a},
\e{609}
where $\ca^{'a}$ and $\pi'_a$ are given by \r{601}, and where $F$
is given by \r{604}.
$\al$ and
$\del$ on the right-hand side  are parameters
defined in \r{216},\r{217}.
The fact that
$\ca^{'a}$,
$\pi'_a$ and their hermitian conjugates
commute  makes the $\del$ in \r{609} satisfy the
properties $\del^2=0$ and $[\del,
\del\dagg]=0$. One may easily check that
$Q=\del+\del\dagg$. The existence of the $\del$-operator \r{609} requires
\r{605}
which is identical to the condition required by the criterion \r{7} for the BRST
singlets which here demands the nonvanishing of \r{603}. Also the existence
of the
oscillators \r{606} requires the same condition. (For nonlinear gauges
there is no
such simple relation as \r{609}.)

Now there are {\em two} ways to solve
$\del|ph\hb=\del\dagg|ph\hb=0$. From \r{609} we have either
\be
&&\ca^{'a}|ph\hb_1=\pi'_a|ph\hb_1=0\quad \Rightarrow\quad|ph\hb_1=e^{[Q,
\psi]}|\tilde{\phi}\hb_1,
\e{610}
or
\be
&&(\ca^{'a})\dagg|ph\hb_1=(\pi'_a)\dagg|ph\hb_1=0\quad
\Rightarrow\quad|ph\hb_2=e^{-[Q,
\psi]}|\tilde{\phi}\hb_1,
\e{611}
where $e^{[Q,
\psi]}$ is given by \r{602}, and
and where $|\tilde{\phi}\hb_1$ satisfies the conditions \r{25}. However,
$|\tilde{\phi}\hb_1$ does not satisfy \r{251}.
 As was explained in \cite{Aux}
conditions like $B|ph\hb=0$,
$B\equiv[Q, C]$ allow for gauge fixing conditions
$C|ph\hb=0$ provided $B$ and $C$
satisfy a closed algebra. By means of gauge
transformations it is always possible to
shift the gauge fixing conditions.
We conclude that the solutions
\r{610} do contain the singlets $|s\hb_1$ in \r{222}, and that \r{611}
contain the
singlets \r{6073}.

For the abelian model we are
considering one implication of
the unitary ambiguity \r{608} is as follows: $\del$
in \r{609} may be  replaced by $\del'=U\del U\dagg$
where $U=e^{i[Q,\tilde{\psi}]}$ where
$\tilde{\psi}$ is odd and hermitian. Thus,
$\del'$ is  a gauge transformed $\del$.
The corresponding solutions of
$\del'|ph\hb'=(\del')\dagg|ph\hb'=0$
is then given by $|ph\hb'_{1,2}=U|ph\hb_{1,2}$
where $|ph\hb_{1,2}$ are the original
solutions in \r{610} and \r{611}. For $U=U_R(\pi/2)$ in \r{39} we
have \eg
\be
&&|ph\hb'_1=e^{-\al K_2-\beta K_1+2\ga K_3+
\del K_4}|\tilde{\phi}\hb_2
\e{612}
where
\be
&&p_a|\tilde{\phi}\hb_2=\bapet^a|\tilde{\phi}\hb_2=0,
\e{613}
which are identical to the conditions on $|\phi\hb_2$ in \r{25}.
The states \r{612} do therefore contain the singlets $|s\hb_2$ in \r{222}.

Although $\del$ and the solutions of \r{2} are defined up to unitary
transformations there are two distinct solutions
of \r{2} which are not connected by
any unitary operator. In the abelian case they
are given by \r{610} and \r{611}. Either there is
an additional condition which excludes one of
these conditions or one has to require
them to be equivalent. The latter option was
used in \eg \cite{Proper}. In the next
section we show that the two solutions
 have opposite norms in some cases.

\setcounter{equation}{0}
\section{Wave function representations}

We consider now wave function representations of
the considered BRST singlets. These
wave functions will be expressed in terms of
the coordinates which are eigenvalues of
$x^a$, $v^a$, $\ca^a$, and $\bapet^a$. Since
the state space as a Fock space contains
half positive and half indefinite metric states
the proposed general rule is that we
can only work with eigenstates which are such
that half of the bosonic and fermionic
coordinates have real eigenvalues and half
imaginary eigenvalues \cite{Proper}. The question
 is
which halves to choose. For the simple
abelian model which we are considering the
results will be that the gauge fixing fermion
$\psi$ to a large extent determines the freedom in this choice.

Some inverses of the unitary
gauge transformations in \r{a5}  appendix A are given by
\be
&&\left(\begin{array}{c}x^a\\v^a\end{array}\right)=
\left(\begin{array}{cc}\tal&\tbeta\\ \tga&\tdel\end{array}\right)
\left(\begin{array}{c}z^a\\w^a\end{array}\right)
\e{701}
\be
&&\left(\begin{array}{c}\ca^a\\ \bapet^a\end{array}\right)=
\left(\begin{array}{cc}\tal&\tbeta\\ \tga&\tdel\end{array}\right)
\left(\begin{array}{c}\eta^a\\ \theta^a\end{array}\right)
\e{702}
where $z^a\equiv x^{'a}$, $w^a\equiv v^{'a}$, $\eta^a\equiv \ca^{'a}$, and
$\theta^a\equiv \bapet^{'a}$. The real constants
$\tal, \tbeta, \tga, \tdel$ may
be obtained from
\r{a5} in appendix A. They satisfy the property
\be
&&\det\left(\begin{array}{cc}\tal&\tbeta\\
\tga&\tdel\end{array}\right)=\tal\tdel-\tga\tbeta=e^{-2d}>0
\e{703}
$z^a, w^a, \eta^a$ and $\theta^a$ are hermitian
operators like $x^a, v^a, \ca^a,
\bapet^a$. We choose $z^a$ and $\eta^a$ to
have real eigenvalues, and $w^a$ and
$\theta^a$ to have imaginary eigenvalues.
The eigenstates  $|z, u\hb$ and
$|\eta, \theta\hb$ satisfy then
\be
&&z^a|z', u\hb=z^{'a}|z', u\hb, \quad w^a|z',
u\hb=iu^a|z', u\hb,\nn\\
&&\eta^a|\eta', \theta'\hb=\eta'|\eta',
\theta'\hb, \quad \theta^a|\eta',
\theta'\hb=i\theta^{'a}|\eta', \theta'\hb.
\e{704}
Note that $|z, u\hb$ are eigenstates to
$x^a$ and $v^a$ with complex eigenvalues.
We have
\be
&&x^a|z, u\hb=(\tal z^a+i\tbeta u^a)|z, u\hb,
\quad v^a|z, u\hb=(\tga z^a+i\tdel
u^a)|z, u\hb.
\e{705}
Similarly  $|\eta, \theta\hb$ are eigenstates
to $\ca^a$ and $\bapet^a$ with complex eigenvalues.
\be
&&\ca^a|\eta, \theta\hb=(\tal \eta^a+i\tbeta
\theta^a)|\eta, \theta\hb, \quad  \bapet^a|\eta,
\theta\hb=(\tga \eta^a+i\tdel
\theta^a)|\eta, \theta\hb
\e{7051}
Note that the state
$|\eta,
\theta\hb$ may be written as
\be
&&|\eta', \theta'\hb=e^{-\eta^{'a}
\pet_{\eta a}-i\theta^{'a}\pet_{\theta
a}}|0\hb_{\eta\theta}
\e{706}
where the vacuum state $|0\hb_{\eta\theta}$
is Grassmann even satisfying the conditions
\be
&&\eta^a|0\hb_{\eta\theta}=\theta^a|0\hb_{\eta\theta}=0.
\e{707}
The states $|z, u\hb$ and $|\eta, \theta\hb$ satisfy also
\be
&&(|z, u\hb)\dagg=\vb z, -u|, \quad
(|\eta, \theta\hb)\dagg=\vb \eta, -\theta|
\e{708}
as well as the completeness relations
\be
&&\int d^nu d^nz |z, u\hb \vb z, u|=\bett, \quad
\int d^n\theta d^n\eta |\eta,
\theta\hb \vb \eta, \theta|=\bett.
\e{709}
If we define the combined eigenstates by
\be
&&|z, u, \eta, \theta\hb\equiv |z,
u\hb\otimes|\eta, \theta\hb,
\e{710}
then we may associate a wave function
to any state $|\psi\hb$ through the relation
\be
&&\psi(z, u, \eta, \theta)\equiv
\vb z, u, \eta, \theta|\psi\hb.
\e{711}
The BRST singlet $|s\hb_1$ satisfy the conditions
\be
&&(x^a+Mv^a)|s\hb_1=0, \quad
(\ca^a+M\bapet^a)|s\hb_1=0, \nn\\
&&(\pi_a-Mp_a)|s\hb_1=0, \quad
(\baca_a-M\pet_a)|s\hb_1=0,
\e{712}
where
\be
&&M\equiv -i{\al B\over A+i\ga B},
\e{713}
where in turn $A$ and $B$ are given by \r{a06}
and \r{a07} in appendix A. The singlet
$|s\hb_2$ satisfies on the other hand
\be
&&(v^a+Nx^a)|s\hb_2=0, \quad  (\bapet^a+N\ca^a)|s\hb_2=0, \nn\\
&&(p_a-N\pi_a)|s\hb_2=0, \quad  (\pet_a-N\baca_a)|s\hb_2=0,
\e{714}
where
\be
&&N\equiv -i{\beta B\over A-i\ga B}.
\e{715}
The solutions of \r{712} and \r{714} are
in the  wave function representation \r{711}
\be
&&\psi_1(z, u, \eta, \theta)=\cA\del^{(n)}
(z+{(\tbeta+M\tdel)\over(\tal+M\tga)}iu)
\del^{(n)}(\eta+{(\tbeta+M\tdel)
\over(\tal+M\tga)}i\theta),\nn\\
&&\psi_2(z, u, \eta, \theta)=\cA'\del^{(n)}
(z+{(\tdel+N\tbeta)\over(\tga+N\tal)}iu)
\del^{(n)}(\eta+{(\tdel+N\tbeta)
\over(\tga+N\tal)}i\theta).
\e{716}
In order for these solutions to make
sense the bosonic delta function must
have a real argument. This
restricts the gauge transformations above, or  in other
words the choice of gauge fixing
fermion $\psi$ governs the quantization rules, \ie it determines
which variables  may be chosen to
have imaginary eigenvalues. These conditions are
explicitly
\be
&&Re\{(\tal+M\tga)(\tbeta+M^*\tdel)\}=0,
\quad Re\{(\tdel+N\tbeta)(\tga+N^*\tal)\}=0.
\e{717}
Note that
\be
&&\sign(Im\{(\tal+M\tga)(\tbeta+M^*\tdel)\})=
-\sign(Im M)=\sign(\al\sin2\sqrt{\al\beta+\ga^2}),\nn\\
&&\sign(Im\{(\tdel+N\tbeta)(\tga+N^*\tal)\})=
-\sign(Im N)=\sign(\beta\sin2\sqrt{\al\beta+\ga^2}),
\e{718}
When calculating the norms of $|s\hb_{1,2}$ we find
\be
&&\int d^nu d^nz d^n\theta d^n\eta \psi^*_1
(z, -u, \eta, -\theta)\psi_1(z, u, \eta,
\theta)\propto (\sign(\al\sin2
\sqrt{\al\beta+\ga^2}))^n, \nn\\
&&\int d^nu d^nz d^n\theta d^n\eta
\psi^*_2(z, -u, \eta, -\theta)\psi_2(z, u, \eta,
\theta)\propto (-\sign(\beta\sin2
\sqrt{\al\beta+\ga^2}))^n,\nn\\
\e{719}
which shows that the norms are undefined
 for exactly those values of $\al, \beta,
\ga$ and $\del$ for which we could not do
the decomposition \r{1} or equivalently for
those values which do not satisfy the criterion
\r{7}. (There is also an infinite
factor present when the sign factors are zero.)
The results
\r{719} also suggest that the physical vacuum
norm changes sign when we cross the
singularity points in odd dimension $n$.
(Such changes for the fermionic vacua was
given in eq.(4.18) in \cite{Fermionic}.)
However, the relations between the bosonic
vacua are not obvious since they are normally not related.

The above results are considerable
generalizations of previous results
which were for $\beta=\ga=0$ \cite{Geza}. One may
notice that when $\ga=0$, in which case
$Re M=Re N=0$, the unitary transformation in
\r{701}-\r{702} may be chosen to be the
identity transformation. This is not the case
when $\ga\neq0$. Note that the conditions
\r{717} leave a three parameter freedom in
the unitary gauge transformations.
Inserting \r{a5} in \r{717} one finds a relation
between the parameters $a$, $b$ and
$c$ ($d$ may be chosen arbitrary).

We conclude that the basic quantization
rule in \cite{Proper} that half of the
bosonic and half of the fermionic unphysical
hermitian operators should have
imaginary eigenvalues leads to perfectly
consistent solutions even in the
general situation which we are considering
here. However,  we had to use this
principle in its most general form in which the
original operators were chosen to have
 complex eigenvalues.

\setcounter{equation}{0}
\section{Still more general forms of $\psi$}
We may consider a still more general gauge
fixing fermion if we view \r{21} as the BRST
charge of a bosonic gauge theory. In this
case only the first ghost number operator $N_b$ in
\r{215} has to be conserved. The most general
invariant quadratic gauge fixing fermion
allowed by the general expression
\r{213} is then
\be
&&\psi'=\psi+\zeta_1\pet_ap^a+\zeta_2\baca_a\pi^a+
\zeta_3\pet_a\pi^a+\zeta_4\baca_ap^a,
\e{41}
where
$\psi$  is given by \r{216}. This expression yields
\be
&&[Q, \psi']=\al K_1+\beta K_2+2\ga K_3+\del K_4
+\la_1V_1+\la_2V_2+\la_3V_3,
\e{42}
where in turn
\be
&&V_1\equiv[Q, \pet_ap^a]=p^ap_a,\quad V_2\equiv-
[Q, \baca_a\pi^a]=-\pi_a\pi^a,\nn\\&&
V_3\equiv[Q, \pet_a\pi^a]=[Q, \baca_ap^a]=p_a\pi^a.
\e{43}
Note that $\pet_a\pi^a=\baca_ap^a+[Q, \pet_a\baca^a]$.
In distinction to the case in the previous
sections the metric $g_{ab}$ enters here
explicitly. In fact, in order to have
quadratic expressions in \r{41} and \r{42} the
metric $g_{ab}$ must be constant, and this
was assumed in \r{43}. Comparison between
\r{41} and \r{42} implies the following
relations between the parameters
$\zeta_i$ and
$\la_k$: $\la_1=\zeta_1$, $\la_2=-\zeta_2$
and $\la_3=\zeta_3+\zeta_4$. If $\del\neq0$ we may choose $\zeta_3-\zeta_4$
such
that $\psi^2=0$ without affecting the gauge fixing factor $e^{[Q, \psi]}$.

Obviously $V_k$ commute among themselves.
Note that they cannot provide for a good
gauge fixing by themselves since they do
not contain ghost terms. The combined algebra
of $K_i$ and $V_k$ is given by \r{219}, \r{220} and
\be
&&[K_1, V_1]=0,\;\;\;[K_1, V_2]=-2iV_3,
\;\;\;[K_1, V_3]=iV_1,\nn\\&&[K_2,
V_1]=2iV_3,\;\;\;[K_2, V_2]=0,\;\;
\;[K_2, V_3]=-iV_2,\nn\\ &&[K_3,
V_1]=-iV_1,\;\;\;[K_3, V_2]=iV_2,
\;\;\;[K_3, V_3]=0,
\e{44}
\be
&&[K_4, V_k]=i V_k\quad k=1,2,3.
\e{45}
Eq.\r{44} implies that $V_k$ transforms
as a vector under the SL(2,R) generated by
$K_1$, $K_2$ and $K_3$, and \r{45} that
$K_4$ generates scale transformations of $V_k$.
The $V_k$-operators satisfy furthermore the relations
\be
&&V_1|\phi\hb_2=0,\quad V_2|\phi\hb_1=0,
\quad V_3|\phi\hb_1=V_3|\phi\hb_2=0.
\e{46}

The general gauge fixing \r{41} involving all
the terms presented so far leads to the
following BRST singlets (see appendix B):
\be
&&|s, \mu\hb_1=e^{[Q, \psi']}|\phi\hb_1=
e^{\al' K_1+\mu p^ap_a}|\phi\hb_1,\nn\\&&
|s, \nu\hb_2=e^{[Q, \psi']}|\phi\hb_2=e^{\beta'
K_2+\nu\pi^a\pi_a}|\phi\hb_2,
\e{47}
$\al'$ and $\beta'$ are still
given by \r{223}-\r{225}, but $\mu$ and
$\nu$ are expressions which also involve the parameters
$\la_i$ in \r{42} (see \r{b16} appendix B). Nonzero $\mu$ or
$\nu$ leads to a regular effective BRST invariant
Hamiltonian or Lagrangian in a bosonic gauge theory.  We note
that the
 duality property \r{27} here
generalizes to
\be
&&e^{\al K_1+\mu p^ap_a}|\phi\hb_1=
e^{-{1\over\al} K_2-{\mu\over\al^2}
\pi^a\pi_a}|\phi\hb_2.
\e{48}
This implies (cf.\r{321})
\be
&&U_S(\ln \al^2)U_R(\pi/2)e^{\al K_1+
\mu p^ap_a}|\phi\hb_1=e^{-{1\over \al}
K_2+{\mu\over\al^2}
\pi^a\pi_a}|\phi\hb_2=e^{\al K_1-
\mu p^ap_a}|\phi\hb_1.
\e{49}
Thus, states with opposite signs of
$\mu$ ($\nu$) are unitarily equivalent. In fact, here we have a much larger
set of
unitary gauge transformations than those in section 3 since we  only have
to require
$U=e^{i[Q, \tpsi']}$ to commute with the first ghost number
operator $N_b$ in \r{215}.
$\tpsi'$ may therefore have the same form as $\psi'$ in \r{41}.
We have now \eg gauge transformations of the form
\be
&&U(p_i)=e^{ip_1V_1+ip_2V_2+ip_3V_3},
\e{4901}
where $p_i$ are real constants,
corresponding to the last terms in \r{41}. In fact, in appendix B it is
shown that
arbitrary values of $\la_i$ in \r{42} may be reached from $\la_i=0$ by unitary
transformations with \r{4901} except if $\del=0$ or $(\del/2)^2=\al\beta+\ga^2$.
($p_i$ are infinite for these values.) We note that
$\del\neq0$ allows for $\psi'$ in
\r{41} to be chosen nilpotent, and that
$(\del/2)^2\neq\al\beta+\ga^2$ is required  from the
classical requirement that $\Lambda^a=0$ and $\chi^a=0$ together with the
original
first class constraints must constitute a set of second class constraints.
We have  the relations
\be
&&U(p_i)e^{\al K_1+\beta K_2+2\ga K_3+\del K_4}|\phi\hb_{1,2}=
e^{\al \tilde{K}_1+\beta
\tilde{K}_2+2\ga \tilde{K}_3+\del
\tilde{K}_4}\tilde{|\phi\hb}_{1,2}, \nn\\&&
\tilde{|\phi\hb}_{1,2}\equiv U(p_i)|\phi\hb_{1,2},
\e{4902}
where $\tilde{K}_i\equiv U(p_i)K_iU\dagg(p_i)$ satisfy the
same algebra as $K_i$. For
$p_1=p_2=0$, $p_3=\la$ we have in particular
\be
&&\tilde{|\phi\hb}_{1,2}=|\phi\hb_{1,2},\quad \tilde{K}_1=K_1+\la p^ap_a,\nn\\
&&\tilde{K}_2=K_2+\la\pi^a\pi_a,\quad
\tilde{K}_3=K_3,\quad \tilde{K}_4=K_4+\la p^a\pi_a.
\e{410}
In this case the reduction formulas in section 2 may be used since
$\tilde{K}_i$  annihilate the same
$|\phi\hb$-states as $K_i$. This corresponds to the parameter values
\be
&&\la_1=\al\la,\quad \la_2=-\beta\la,\quad \la_3=\del\la
\e{411}
in \r{42}.
We find here
\be
&&U(\la)|s,0\hb_1=|s, \al'\la\hb_1, \quad U(\la)|s,0\hb_2=|s,
\beta'\la\hb_2,\quad
U(\la)\equiv e^{i\la V_3}
\e{412}
in terms of the states \r{47}. Hence, the parameters $\mu$ and $\nu$ in
\r{47} are
pure gauge parameters.

A still more general gauge fixing fermion of the general form \r{213} is
\be
&&\psi=\pet_a\Lambda^a+\baca_a\chi^a,
\e{4121}
where
\be
&&\Lambda^a=\al v^a+(\del/2-\ga)x^a+\zeta_1 p^a+\zeta_3 \pi^a-\Lambda^a_0, \nn\\
&&\chi^a=\beta x^a+(\del/2+\ga)v^a+\zeta_2 \pi^a+\zeta_5 p^a-\chi^a_0,
\e{4122}
where in turn $\Lambda^a_0$ and $\chi^a_0$ are constants. (For
$\Lambda^a_0=0$ and
$\chi^a_0=0$
\r{4121} agrees with
\r{41}.) The corresponding inner product states $e^{[Q, \psi]}|\phi\hb$ may be
obtained by means of unitary gauge transformations of the form ($x^a_0$ and
$v^a_0$
are real constants)
\be
&&U=e^{i[Q, \pet_a x^a_0+\baca_a v^a_0]}=e^{ix^a_0 p_a+iv^a_0 \pi_a}
\e{4123}
provided $(\del/2)^2\neq\al\beta+\ga^2$. This condition we recognize as the
condition for  $\Lambda^a$ and
$\chi^a$ to be consistent gauge fixing variables classically. If
$(\del/2)^2=\al\beta+\ga^2$ then the inner product states from \r{41} and
\r{4121}
seems to be gauge inequivalent although  the corresponding inner product
states
\r{47} are  well defined at these values.

From appendix B it follows that the properties
\r{603}-\r{605} as well as \r{610} are still valid for the gauge fixing fermion
\r{41}.  Thus, the existence conditions of
BRST singlets as inner product states are
exactly the same as before (see section 4) for
arbitrary values of $\la_i$ in \r{42}.
However, the wave function representations
are different from those in section 5. The
singlets $|s\hb_{1,2}$ from \r{41} are here determined
by the conditions \r{712} and \r{714} in
which the first conditions are replaced by
\be
&&x^{'a}|s\hb_1=(A(x^a+Mv^a)+Rp^a+S\pi^a)|s\hb_1=0,\nn\\
&&v^{'a}|s\hb_2=(A'(v^a+Nx^a)+R'p^a+S'\pi^a)|s\hb_2=0,
\e{491}
where  $x^{'a}$ and $v^{'a}$ are given in \r{b7}
appendix B. (The complex constants $A, R, S$, and $A', R', S'$ are obtained
from \r{b7}, \r{b8} and \r{a2}.)    The solutions are of the form
\be
&&\psi_1(z, u, \eta, \theta)=\cA e^{K(\tal+M\tga)^2
\{z^a+{(\tbeta+M\tdel)\over(\tal+M\tga)}iu^a\}^2}
\times\del^{(n)}(\eta+{(\tbeta+M\tdel)\over(\tal+M\tga)}i\theta),\nn\\
&&\psi_2(z, u, \eta, \theta)=\cA'
e^{K'(\tga+N\tal)^2\{z^a+{(\tdel+N\tbeta)
\over(\tga+N\tal)}iu^a\}^2}\times\del^{(n)}
(\eta+{(\tdel+N\tbeta)\over(\tga+N\tal)}i\theta),
\e{492}
where
\be
&&K\equiv{A\over 2i(R+SM)}, \quad K'\equiv{A'\over 2i(R'+S'N)}.
\e{493}
If we still impose conditions \r{718} we
find for the norms of $|s\hb_{1,2}$
\be
&&\int d^nu d^nz d^n\theta d^n\eta
\psi^*_1(z, -u, \eta, -\theta)\psi_1(z, u, \eta,
\theta)\propto\nn\\
&&\propto \int d^nu' d^n z'
e^{K(\tal+M\tga)^2z^{'2}+K^*(\tal+M^*\tga)^2u^{'2}}
(\sign(\al\sin2\sqrt{\al\beta+\ga^2}))^n,
\nn\\ &&\int d^nu d^nz d^n\theta d^n\eta
\psi^*_2(z, -u, \eta, -\theta)\psi_2(z, u,
\eta,
\theta)\propto\nn\\
&&\propto \int d^nu' d^n z' e^{K'(\tga+N\tal)^2
z^{'2}+K^{'*}(\tga+M^*\tal)^2u^{'2}}(-
\sign(\beta\sin2\sqrt{\al\beta+\ga^2}))^n.
\e{494}
This is well defined for $\al\sin2\sqrt{\al\beta+\ga^2}
\neq0$ ($\beta\sin2\sqrt{\al\beta+\ga^2}\neq0$)
provided $K(\tal+M\tga)^2$
($K'(\tga+N\tal)^2$) has a negative real part.
This seems always possible to achieve
with an appropriate choice of $\tal$ and $\tga$.
 However, for
$\al\sin2\sqrt{\al\beta+\ga^2}=0$
($\beta\sin2\sqrt{\al\beta+\ga^2}=0$) the
normalizations are undefined since the bosonic
Jacobi determinant is infinite while
the fermionic integration yields zero. Although we have found difficulties with
certain gauge transformations at the values $\del=0$ and
$(\del/2)^2=\al\beta+\ga^2$
the inner products are perfectly well defined at these values.

Above we have investigated some applications of a general gauge
fixing fermion
\be
&&\psi=\pet_a\Lambda^a+\baca_a\chi^a,
\e{413}
where $\Lambda^a$ and $\chi^a$ are linear
and covariant in the basic variables for a
simple abelian bosonic gauge theory. We have
then shown that the properties of the BRST
singlets expressed in terms of $e^{[Q, \psi]}$
in general are determined by the
SL(2,R) properties of the commutators
$[Q,\psi]$. A characteristic feature of these
cases is also that the gauge fixing variables
$\Lambda^a$ and $\chi^a$ may in general be chosen
to commute without affecting $[Q,
\psi]$  which in turn implies that
$\psi$ in \r{413}  may be chosen to be  nilpotent. (Here this was true if
$\del\neq0$.) These properties are expected to be valid
also in more general gauge theories, which
to some extent is verified in the next section.

If we instead  interpret $Q$ as the BRST
charge of a fermionic gauge theory then
the general form \r{214} allows for the gauge
fixing fermion
\be
&&\psi'=\psi+\rho_1\ca_ax^a+\rho_2
\bapet_a x^a+\rho_3\ca_a v^a+\rho_4\bapet_av^a,
\e{414}
where
$\psi$  is given by \r{216}. This
expression yields in the case of a symmetric metric
\be
&&[Q, \psi']=\al K_1+\beta K_2+2\ga
K_3+\del K_4+i\rho \bapet_a\ca^a,
\e{415}
where in turn $\rho=\rho_2-\rho_3$. Note that
\be
&&[Q, \ca_ax^a]=[Q,\bapet_av^a]=0, \nn\\&& [Q, \bapet_a x^a]=- [Q, \ca_a v^a
]=i\bapet_a\ca^a.
\e{416}
Note also that $K_i$
commute with $\bapet_a\ca^a$, but that
\be
&&[K_4, \bapet_a\ca^a]=-i\bapet_a\ca^a,
\e{417}
which means that $\bapet_a\ca^a$ is an SL(2,R) scalar that scales under $K_4$.
$\psi'$ in \r{414}  may be chosen to be nilpotent without affecting $[Q, \psi']$
provided $\del\neq0$ {\em and} $(\del/2)^2\neq\al\beta+\ga^2$. The last
condition
follows also if we require $\tilde{\chi}_a$ and
$\tilde{\Lambda}_a$ in \r{214} to be
consistent gauge fixing variables. Also here we have a much larger set of
unitary gauge transformations than those in section 3 since we  only have
to require
$U=e^{i[Q, \tpsi']}$ to commute with the second ghost number operator $N_f$ in
\r{215}.
$\tpsi'$ may therefore have the same form as $\psi'$ in \r{414}.
We have now  gauge transformations of the form
\be
&&U(\la)=e^{\la\bapet_a\ca^a},
\e{418}
where $\la$ is a real constant,
corresponding to the last terms in \r{415}. We find from \r{418}
\be
&&U(\la)e^{\al K_1+\beta K_2+2\ga K_3+\del K_4}|\phi\hb_{1,2}=
e^{\al \tilde{K}_1+\beta
\tilde{K}_2+2\ga \tilde{K}_3+\del
\tilde{K}_4}{|\phi\hb}_{1,2}=\nn\\&&=e^{\al K_1+\beta K_2+2\ga
K_3+\del K_4+i\del\la \bapet_a\ca^a}{|\phi\hb}_{1,2},
\e{419}
since $\tilde{K}_i\equiv U(\la)K_iU\dagg(\la)=K_i$ and $\tilde{K}_4\equiv
U(\la)K_4U\dagg(\la)=K_4+i\la\bapet_a\ca^a$. Hence, arbitrary values of
$\rho$ in
\r{416} may be reached from the gauge fixing $\psi$
provided $\del\neq0$ a condition
which also the nilpotence of $\psi'$ in \r{414} required.

\setcounter{equation}{0}
\section{Applications to nontrivial models}
In this section we give two  examples
of non-trivial bosonic gauge theories to
which our formulas may easily be applied.
\subsection{Example 1: Cohomological dynamics}
Any regular dynamical system may be
written in a reparametrization invariant form
\cite{Lan}. The BFV-BRST charge for such a theory is
\be
&&Q=\ca(\pi+H)+\bapet\pi_v,
\e{51}
where $H$ is the Hamiltonian of the
original theory and $\pi$ the conjugate momentum
to time, $t$,  which here is a dynamical
variable. $\pi_v$ is a conjugate momentum to
a Lagrange multiplier $v$.
All variables are hermitian.
Since $H$ commutes with $t$ we may treat
$\pi+H$ as conjugate momentum to $t$. It is
then quite obvious that all formulas in
their most general form given in section 6
and appendix B apply here. The BRST
singlets are of the form
\be
&&|s\hb=e^{[Q, \psi']}|\phi\hb,
\e{52}
where $\psi'$ and $[Q, \psi']$ are given
by \r{41}-\r{43}. The $|\phi\hb$-state may
\eg be chosen to be $|\phi\hb_1$ or
$|\phi\hb_2$ determined by the conditions
\be
&&t|\phi\hb_1=\ca|\phi\hb_1=
\baca|\phi\hb_1=\pi_v|\phi\hb_1=0,
\e{53}
\be
&&v|\phi\hb_2=\bapet|\phi\hb_2=
\pet|\phi\hb_2=(\pi+H)|\phi\hb_2=0.
\e{54}
Note that the last condition is the
Schr\"{o}dinger equation. In this case we have no
covariant indices to care about.
We may therefore without problems consider
transformations obtained from the BRST invariant operator
$V$ given by (cf \r{4123})
\be
&&V\equiv e^{i[Q, t_0\pet+v_0\baca]},
\e{55}
where $t_0$ and $v_0$ are two constants. We find then
\be
&&V|s\hb=e^{[Q, \tilde{\psi}]}|\tilde{\phi}
\hb,\quad \tilde{\psi}=V\psi' V^{-1},\quad
|\tilde{\phi}\hb=V|\phi\hb.
\e{56}
 $\tilde{\psi}$ is equal to $\psi'$ with
$t$ and $v$ replaced by
$(t-t_0)$ and
$(v-v_0)$, and  $|\tilde{\phi}\hb_{1,2}$
satisfy \r{53} and \r{54} with the first
conditions replaced by
\be
&&(t-t_0)|\tilde{\phi}\hb_{1}=0    ,
\quad (v-v_0)|\tilde{\phi}\hb_{2}=0.
\e{58}
We have then arrived at the gauge
fixings considered in \cite{Time}. However, the
singlets $V|s\hb$ are a considerable
generalization of those given in \cite{Time}.

\subsection{Example 2: Nonabelian gauge theory}
The BFV-BRST charge for a general
bosonic nonabelian gauge theory with a finite number
of degrees of freedom is given by
($a, b, c =1,\ldots,n<\infty$)\cite{BFV}
\be
&&Q=\theta_a\ca^a-\frac{1}{2}iU_{bc}^{\;\;\;a}\pet_a
\ca^b\ca^c-\frac{1}{2}iU_{ab}^{\;\;\;b}\ca^a + \bapet^a\pi_a
\e{510}
where $\theta_a$ are the hermitian bosonic gauge generators
(constraints) satisfying
\be
&[\theta_a, \theta_b]_{-}=iU_{ab}^{\;\;\;c}\theta_c
\e{511}
where $U_{ab}^{\;\;\;c}$ are  real
structure constants. (We consider Lie group
theories.) Remarkably enough there exists
a simple abelianization of the BRST charge
\r{510} by means of which all our results
for abelian models  can be directly applied
to
\r{510} \cite{Geza} (see also \cite{SH}).
Introduce canonical group coordinates,
$x^a$, which are hermitian operators satisfying the properties
\be
&&[x^a, x^b]=0, \quad [x^a, \theta_b]=iM^a_{\;\;b}(x),
\e{512}
where the hermitian matrix operators $M^a_{\;\;b}(x)$ satisfy
$M^a_{\;\;b}(x)x^b=x^a$, $M^a_{\;\;b}(0)
=\del^a_b$ and the Maurer-Cartan equations
\be
&&(\partial_d
M^c_{\;\;a})M^d_{\;\;b}-(\partial_d
M^c_{\;\;b})M^d_{\;\;a}=U_{ab}^{\;\;\;d}M^c_{\;\;d}.
\e{513}
We may then define hermitian conjugate
momenta, $p_a$, to $x^a$ by
\be
&&p_a\equiv(M^{-1})^b_{\;\;a}(x)\theta_b+i\halv
(M^{-1})^b_{\;\;a}(x)\partial_cM^c_{\;\;b}(x)
\e{514}
in terms of which we have
\be
&&\theta_a=\halv\left(p_bM^b_{\;\;a}(x) +
 M^b_{\;\;a}(x)p_b  \right).
\e{515}
Consider then also the following unitary transformation
which only affects $\ca^a$, $\pet_a$,
and $p_a$
\be
&&\ca^a\;\ra\;\tca^a=M^a_{\;\;b}(x)\ca^b,
\;\;\;\pet_a\;\ra\;\tpet_a=(M^{-1})^b_{\;\;a}(x)\pet_b\nn\\
&&p_a\;\ra\;\tpe_a=p_a+i\halv
\partial_aM^b_{\;\;c}(x)(M^{-1})^d_{\;\;b}
(\ca^c\pet_d-\pet_d\ca^c).
 \e{516}
If one now inserts \r{515} into \r{510}
and replaces $\ca^a$, $\pet_a$, and $p_a$ by
$\tca^a$, $\tpet_a$, and $\tpe_a$ using \r{516}, one finds
\be
&&Q=\tca^a\tpe_a+\pi_a\bapet^a,
\e{517}
which is the BRST charge \r{21}
for an abelian model. The complete set of
canonical operators are $\{x^a,
\tpe_a; v^a, \pi_a; \tca^a, \tpet_a; \baca_a,
\bapet^a\}$. Since we here have a
bosonic gauge theory we may consider the general
gauge fixing fermion
\r{41}. The BRST singlets are
given by expressions of the form
\be
&&|s\hb=e^{[Q, \psi']}|\phi\hb,
\e{518}
where $[Q, \psi']$ is given by
\r{42}. The  $K_i$-operators \r{28} and \r{218} may
then be written in the
following invariant forms
\be
&K_1\equiv&[Q, \tpet_av^a]=[Q,
\pet_a(M^{-1})^a_{\;\;b}v^b]=\theta_a(M^{-1})^a_{\;\;b}v^b-
i\half U_{bc}^{\;\;\;c}(M^{-1})^b_{\;\;a}  v^a-\nn\\&&-
i\dif_aM^b_{\;\;c}(M^{-1})^d_{\;\;b}\pet_d\ca^cv^a+
i\pet_a(M^{-1})^a_{\;\;b}\bapet^b,\nn\\
&K_2\equiv&[Q,
\baca_a
x^a]=x^a \pi_a+i\baca_a  M^a_{\;\;b}\ca^b,\nn\\
&K_3\equiv&\half[Q, \baca_av^a-\tpet_ax^a]=\half[Q,
\baca_av^a-\pet_ax^a]=\half\left(v^a\pi_a-\theta_ax^a+\half
iU_{ab}^{\;\;\;b}x^a-\right.\nn\\&&\left.- i x^a
U^{\;\;\;d}_{ab}\pet_d\ca^b-
i\pet_aM^a_{\;\;b}\ca^b-i\bapet^a\baca_a\right)=
\half\biggl(\pi_av^a-x^a\theta_a-
\biggr.\nn\\&&\left.-\half
iU_{ab}^{\;\;\;b}x^a+i x^aU^{\;\;\;d}_{ab}\ca^b\pet_d+
iM^a_{\;\;b}\ca^b\pet_a+i\baca_a\bapet^a\right),\nn\\
&K_4\equiv&\half[Q, \baca_av^a+\tpet_ax^a]=\half[Q,
\baca_av^a+\pet_ax^a]=\half\left(v^a\pi_a+\theta_ax^a-\half
iU_{ab}^{\;\;\;b}x^a+\right.\nn\\&&\left.+ i
x^aU^{\;\;\;d}_{ab}\pet_d\ca^b+
i\pet_aM^a_{\;\;b}\ca^b-i\bapet^a\baca_a\right)
=\half\biggl(\pi_av^a+x^a\theta_a+
\biggr.\nn\\&&\left.+\half
iU_{ab}^{\;\;\;b}x^a-i x^aU^{\;\;\;d}_{ab}\ca^b\pet_d-
iM^b_{\;\;a}\ca^a\pet_b+i\baca_a\bapet^a\right),
\e{519}
However, the $V_i$-operators \r{43} are noncovariant. The reason is that
\r{41}-\r{43} require $\tpe^a$ to be
obtained from $\tpe_a$ by means of a constant
metric $g^{ab}$, which is unnatural
since $\tpe_a$ has a curved index.

The state $|\phi\hb$ in \r{518} may
\eg be chosen to be $|\phi\hb_1$ or
$|\phi\hb_2$  determined by the conditions
\be
&&x^a|\phi\hb_1=\tca^a|\phi\hb_1=
\baca_a|\phi\hb_1=\pi_a|\phi\hb_1=0,\nn\\
&& v^a|\phi\hb_2=\bapet^a|\phi\hb_2=
\tpet_a|\phi\hb_2=\tpe_a|\phi\hb_2=0.
\e{520}
 In terms of the original variables these conditions are
\be
&&x^a|\phi\hb_1=\ca^a|\phi\hb_1=
\baca_a|\phi\hb_1=\pi_a|\phi\hb_1=0,\nn\\
&& v^a|\phi\hb_2=\bapet^a|\phi\hb_2=
\pet_a|\phi\hb_2=\left(\theta_a+i\halv
U_{ab}^{\;\;\;b}\right)
 |\phi\hb_2=0.
\e{521}
To prove the equivalence of the last
conditions in \r{520} and \r{521} we have used
\be
&&(M^{-1})^b_{\;\;a}\dif_cM^c_{\;\;b}-
(M^{-1})^b_{\;\;c}\dif_aM^c_{\;\;b}
=(M^{-1})^b_{\;\;a}U_{bc}^{\;\;\;c},
\e{522}
which follows from \r{513}.
 Concerning the last condition in \r{521} one may note that
\be
&&[Q, \pet_a]|\phi\hb_2=\left(\theta_a+
i\halv U_{ab}^{\;\;\;b} \right)|\phi\hb_2.
\e{523}
Thus, $|\phi\hb_1$ and
$|\phi\hb_2$ are determined by manifestly
covariant conditions. Note also the
invariant properties \r{221}.  If furthermore
the regulator factor $e^{[Q, \psi]}$
is of the general form considered in sections
2-5 and appendix A, then also the
singlets $|s\hb_{1,2}=e^{[Q, \psi]}
|\phi\hb_{1,2}$ satisfy manifestly covariant
conditions which may be extracted from
\r{a2}. (The special formulas (6.19)-(6.20)
in \cite{Geza} are not quite correct. The
adjoint matrix representation $A^a_{\;\;b}$
in (6.19) should have the argument
$+i\al' v^a$, and the right-hand side of
the last equality in (6.20) should have a
term $i\beta' M^b_{\;\;a}\pi_b$ added.)
The singlets \r{518} are a considerable
generalization of those given in \cite{Geza}.

\setcounter{equation}{0}
\section{Conclusions}

We have made an extensive analysis  of the abstract operator
approaches to BRST quantization on inner product spaces as presented in the
introduction. For simple abelian models which allow
for explicit calculations we have
verified the properties and interrelations described
in the introduction. For these
models we have considered the most general gauge
fixing fermions expressed in terms of quadratic
invariant terms, and then determined
the exact conditions for  the existence of BRST singlets as inner
product states both abstractly and concretely
within a specific state representation. The
 abstract conditions as well as the
explicit wave function integrations led to
the same results.  We have verified the
general quantization  rule that half of the fundamental hermitian
operators are to be quantized with imaginary
eigenvalues. The choices of gauge fixing fermions
$\psi$ were shown to partly govern these quantization rules.
Some choices of
$\psi$ forced us even  to consider
complex eigenvalues of the basic variables. In this way we have
 to some extent explored the freedom this general quantization
rule leaves us. We have also found that the  importance of the
$SL(2,R)$ symmetry in the gauge fixing factor to determine the BRST singlets,
first noted in \cite{Geza}, was retained even in
the much more general invariant gauge fixing  fermions
considered here. If we also introduce parameters in $\psi$ then this $SL(2,R)$
symmetry will be broken unless these parameters are introduced by means of
unitary
gauge transformations as was done in section 6 and at the end of subsection 7.1.
We have also investigated unitary gauge transformations and shown that there are
natural extended ones performing SL(2,R) transformations although they
effectively
only act as abelian gauge transformations on the BRST invariant states.
Conditions
for gauge equivalences as well as whether or not $\psi$ may be chosen nilpotent
without affecting the gauge fixing factor were determined.
Although our treatment only is valid for simple abelian models the results
should be
valid in general since we always may perform  an abelianization within a
sector of
a given model by an appropriate choice of $\psi$. In subsection 7.2 this was
demonstrated for a general nonabelian model. Finally,
it should be mentioned that the
corresponding path integral results for the models considered here may be
directly
extracted from our treatment.

\setcounter{section}{1}
\setcounter{equation}{0}
\renewcommand{\thesection}{\Alph{section}}
\newpage
\noindent
{\Large{\bf{Appendix A}}}
 \vspace{5mm}\\
{\bf Transformation formulas used in sections 2 and 3}\\ \\
In section 2 and 3 we considered the general
 gauge fixing fermion \r{216}. It leads to
the regulator factor
\be
&&e^{[Q,\psi]}= e^{\al K_1+\beta K_2+2\ga K_3+\del K_4},
\e{a01}
where $\al, \beta, \ga$ and $\del$ are real
 parameters. By means of the commutation
relations
\r{219}-\r{220} of the $K_i$-operators we may
factorize this regulator factor in
various way. We have \eg the following alternative forms
\be
&&e^{[Q,\psi]}=
e^{is_1 K_1}e^{s_2 K_2}e^{is_3 K_1}e^{\delta K_4},\quad \beta\neq0
\e{a02}
\be
&&e^{[Q,\psi]}=
e^{it_1 K_2}e^{t_2 K_1}e^{it_3 K_2}e^{\delta K_4},\quad \al\neq0
\e{a03}
where
\be
&&s_1  =  \frac{i(A-i\gamma B-1)}{\beta B}, \quad
s_2  =  \beta B, \quad
s_3  = \frac{i(A+i\gamma B-1)}{\beta B},
\e{a04}
and
\be
&&t_1  =\frac{i (A-i\gamma B-1)}{\alpha B}, \quad
t_2  = \alpha B, \quad
t_3  =\frac{i(A+i\gamma B-1)}{\alpha B},
\e{a05}
where in turn
\be
&&A\equiv\cos\sqrt{\al\beta+\ga^2},\quad
B\equiv\frac{\sin{\sqrt{\al\beta+\ga^2}}}{\sqrt{\al\beta+\ga^2}}.
\e{a06}
for $\al\beta+\ga^2>0$. For $\al\beta+\ga^2<0$ we have
\be
&&A\equiv\cosh{\sqrt{-\al\beta-\ga^2}},\quad
B\equiv\frac{\sinh{\sqrt{-\al\beta-\ga^2}}}{\sqrt{-\al\beta-\ga^2}}.
\e{a07}
($A=1$ and $B=1$ for $\al\beta+\ga^2=0$.)

We have used the factorizations above
to derive  transformations of the form
\be
&&D'=e^{[Q,\psi]}De^{-[Q,\psi]} =
e^{\al K_1+\beta K_2+2\ga K_3+\del K_4}D
e^{-\al K_1-\beta K_2-2\ga K_3-\del K_4},
\e{a1}
where $D$ is any operator. For the basic variables we find explicitly
\be
&&{x'}^a=\left(x^a (A+i\ga B)-i\al v^aB
\right)e^{-i\frac{\del}{2}},\nn\\
&&v'^a= \left(v^a(A-i\ga B) -i\beta x^a
B\right)e^{-i\frac{\del}{2}},\nn\\
&&{\ca'}^a=\left(\ca^a(A+i\ga B)-i\al\bapet^a
B\right)e^{-i\frac{\del}{2}},\nn\\
&&{\bapet}^{'a}=\left(\bapet^a(A-i\ga B)-i\beta
\ca^aB\right)e^{-i\frac{\del}{2}},\nn\\
&&{\baca'}_a=\left(\baca_a(A+i\ga B)+i\al\pet_a
B\right
)e^{i\frac{\del}{2}},\nn\\
 &&{\pet'}_a=\left(\pet_a(A-i\ga B)+i\beta\baca_a
B\right)e^{i\frac{\del}{2}},\nn\\
&&{\pi'}_a=\left(\pi_a(A+i\ga B)+i\al
p_aB\right)e^{i\frac{\del}{2}},\nn\\
&&{p'}_a=\left(p_a(A-i\ga B)+i\beta\pi_a
B\right)e^{i\frac{\del}{2}},
\e{a2}
which are well defined even at $\al=0$ and/or $\beta=0$.

These transformations imply
\be
&K_1'=&(A-i\ga B)^2K_1+
\beta^2B^2K_2+2i\beta(A-i\ga B)B
K_3\nn\\
&K_2'=&(A+i\ga B)^2K_2+
\al^2B^2K_1-2i\al(A+i\ga B)B
K_3\nn\\
&K_3'=&(1-2\al\beta B^2)K_3+
i\al(A-i\ga B)BK_1
-i\beta(A+i\ga B)BK_2\nn\\
&K'_4=&K_4
\e{a3}

The corresponding transformations
for the unitary SL(2,R)$\times$U(1)
transformations considered in section 3 are
\be
&&D'=e^{iaK_1+ibK_2+2icK_3+idK_4}De^{-iaK_1-ibK_2-2icK_3-idK_4}.
\e{a4}
The explicit expressions are obtained
from \r{a2} with the replacements $\al\ra ia$,
$\beta\ra ib$,
$\ga\ra ic$ and
$\del\ra id$. We find
\be
&&{x'}^a=\left(x^a (A-cB)+a v^aB\right)e^{\frac{d}{2}},\nn\\
&&v'^a= \left(v^a(A+cB) +b x^a
B\right)e^{\frac{d}{2}},\nn\\
&&{\ca'}^a=\left(\ca^a(A-cB)+a\bapet^a
B\right)e^{\frac{d}{2}},\nn\\
&&{\bapet}^{'a}=\left(\bapet^a(A+cB)+b
\ca^aB\right)e^{\frac{d}{2}},\nn\\
&&{\baca'}_a=\left(\baca_a(A-cB)-a\pet_a
B\right
)e^{-\frac{d}{2}},\nn\\
 &&{\pet'}_a=\left(\pet_a(A+cB)-b\baca_a
B\right)e^{-\frac{d}{2}},\nn\\
&&{\pi'}_a=\left(\pi_a(A-cB)-a
p_aB\right)e^{-\frac{d}{2}},\nn\\
&&{p'}_a=\left(p_a(A+cB)-b\pi_a
B\right)e^{-\frac{d}{2}},
\e{a5}
where now
\be
&&A\equiv\cosh\sqrt{ab+c^2},\quad B\equiv
\frac{\sinh{\sqrt{ab+c^2}}}{\sqrt{ab+c^2}}
\e{a6}
for $ab+c^2>0$. For $ab+c^2<0$ we have to use the replacement
\be
&&\cosh{\sqrt{ab+c^2}}\,\ra\,\cos{\sqrt{-ab-c^2}},\;\;\;
\frac{\sinh{\sqrt{ab+c^2}}}{\sqrt{ab+c^2}}\,\ra\,
\frac{\sin{\sqrt{-ab-c^2}}}{\sqrt{-ab-c^2}}.
\e{a7}
The unitary transformations \r{a5} imply
\be
&K_1'=&(A+cB)^2K_1-
b^2B^2K_2-2b(A+cB)B
K_3\nn\\
&K_2'=&(A-cB)^2K_2-
a^2B^2K_1+2a(A-cB)B
K_3\nn\\
&K_3'=&(1+2abB^2)K_3-
a(A+cB)BK_1
+b(A-cB)BK_2\nn\\
&K'_4=&K_4
\e{a8}
\newpage
\noindent
{\bf Derivation of the reduction formulas \r{222}.}\\ \\
In order to calculate the regulator factor \r{a01} on $|\phi\hb$-states as in
\r{222} the following factorizations are convenient to use
\be
&&e^{[Q,\psi]}=
e^{f_1 K_1}e^{(f_2+if_3)K_3}e^{f_4 K_2}e^{\delta K_4},
\e{a9}
\be
&&e^{[Q,\psi]}=
e^{g_1 K_2}e^{(g_2+ig_3)K_3}e^{g_4 K_1}
e^{\delta K_4},
\e{a10}
where
\be
&&f_1  =   \frac{\alpha B}{A+i\gamma B},\quad
f_2  =  -\arctan{(\frac{\gamma B}{A})},\nn\\
&&f_3  =  \ln|A+i\gamma B|,\quad
f_4  =   \frac{\beta B}{A+i\gamma B},
\e{a11}
\be
&&g_1  =   \frac{\beta B}{A-i\gamma B},\quad
g_2  =  -\arctan{(\frac{\gamma B}{A})},\nn\\
&&g_3  =  -\ln|A+i\gamma B|,\quad
g_4  =  \frac{\alpha B}{A-i\gamma B}.
\e{a12}\\ \\
\setcounter{section}{2}
\setcounter{equation}{0}
\noindent
{\Large{\bf{Appendix B}}}
 \vspace{5mm}\\
{\bf Transformation formulas used in section 6}\\ \\
Here we explore the extra features of the theory
that follows from the more general
choice of  gauge fixing fermion  made in
\r{41}, which is valid for simple
abelian bosonic gauge theories. As we will see all
 the relevant properties of the
theory is still dictated by the
$SL(2,{ R})$-sector of the gauge
fixing considered in appendix A. The regulator
factor is here given by
\be
\exp{\{[Q,\psi]\}}=\exp{\{\alpha K_1+
\beta K_2+2 \gamma K_3+\delta K_4+\la_1
V_1+\la_2 V_2+\la_3 V_3\}},
\e{b1}
where $\la_1, \la_2$ and $\la_3$ are new real parameters.
It may \eg be factorized in the
following two forms
\be
&&e^{[Q,\psi]}=
e^{ip_3 V_3}e^{ip_1 V_1}e^{ip_2 V_2}
e^{is_1 K_1}e^{s_2 K_2}e^{is_3 K_1}
e^{\delta K_4}e^{-ip_2 V_2}e^{-ip_1 V_1}e^{-ip_3 V_3},
\e{b2}
\be
&&e^{[Q,\psi]}=e^{ip_3 V_3}e^{ip_1 V_1}e^{ip_2 V_2}
e^{it_1 K_2}e^{t_2 K_1}e^{it_3 K_2}
e^{\delta K_4}e^{-ip_2 V_2}e^{-ip_1 V_1}e^{-ip_3 V_3},
\e{b3}
where the parameters $s_i$ and $t_i$ are
given in \r{a04} and \r{a05}, and where
\be
p_1=\frac{\left(2{(\frac{\delta}{2})}^2-\al\beta+
\del\ga\right)\la_1-\al^2\la_2-\al\left(\frac{\delta}{2}+\ga\right)\la_3}
{2{\delta}({(\frac{\delta}{2})}^2-\alpha\beta-{\gamma}^2)},
\e{b4}
\be
p_2=\frac{-\beta^2\la_1+\left(2(\frac{\delta}{2})^2-\al\beta-
\del\ga\right)\la_2+\beta\left(\frac{\delta}{2}-\ga\right)\la_3}
{2{\delta}({(\frac{\delta}{2})}^2-\alpha\beta-{\gamma}^2)},
\e{b5}
\be
p_3=\frac{-\beta\left(\frac{\delta}{2}+\ga\right)\la_1+\al\left(\frac{\delta
}{2}-
\ga\right)\la_2+\left((\frac{\delta}{2})^2-\ga^2\right)\la_3}
{{\delta}
({(\frac{\delta}{2})}^2-\alpha\beta-{\gamma}^2)}.
\e{b6}
These results imply that
\be
&&e^{[Q,\psi']}=U(p_i)e^{[Q,\psi]}U\dagg(p_i),
\e{b61}
where
\be
&&U(p_i)\equiv e^{ip_1V_1+ip_2V_2+ip_3V_3}
\e{b62}
is a unitary operator for gauge transformations. Thus, the regulator factor
\r{b1}
may be obtained from the one in appendix A by means of a unitary gauge
transformation. However,   note that the parameters $p_i$ in \r{b4}-\r{b6} are
infinite at
$\del\ra0$ and at
$\del\ra\pm2\sqrt{\al\beta+\ga^2}$. (As was mentioned in section 6,
$\del\neq\pm2\sqrt{\al\beta+\ga^2}$ was required for $\Lambda^a$ and $\chi^a$ in
\r{213} to be consistent gauge fixing conditions
classically, and $\del\neq0$ allows
us to choose
$\psi'$ in \r{41} to be nilpotent without affecting the gauge fixing factor
$e^{[Q,\psi']}$.)

 By means of
\r{b3} we find now the transformation formulas  \r{a2}
in appendix A where ${x'}^{a}$
and
${v'}^{a}$ are generalized
 to
\be
&&{x'}^{a}=(x^{a}(A+i\gamma B)-i\alpha B
v^{a})e^{-\frac{i\delta}{2}}+R_1p^{a}+R_2{\pi}^a,\nn\\&&
{v'}^{a}=(v^{a}(A-i\gamma B)-i\beta Bx^a)
e^{-\frac{i\delta}{2}}+R_3p^{a}  +R_4{\pi}^a,
\e{b7}
 where
\be
&&R_{1}\equiv4i p_1 Im(e^{-
\frac{i\delta}{2}}(A+i\gamma B))-2 i \alpha B p_3
\cos{\frac{\delta}{2}},\nn\\
&&R_{2}\equiv-2i(A+i\gamma
B)p_3\sin{\frac{\delta}{2}}+2iB
(e^{-\frac{i\delta}{2}}p_2\alpha-e^{\frac{i\delta}{2}}p_1\beta),\nn\\
&&R_{3}\equiv
-2i(A-i\gamma B)p_3\sin{\frac{\delta}{2}}+2i B({e^{\frac{i\delta}{2}}
p_2\alpha-e^{-\frac{i\delta}{2}}}p_1\beta),\nn\\
&&R_{4}\equiv
4ip_2 Im(e^{\frac{i\delta}{2}}
(A+i\gamma B))-2i\beta B p_3 \cos{\frac{\delta}{2}},
\e{b8}
where the parameters $p_i$ are given by \r{b4}-\r{b6}. The
parameters $R_i$, and therefore also $x^{'a}$ and $v^{'a}$,  are perfectly well
defined and finite at
$\del\ra0$,
$\del\ra\pm2\sqrt{\al\beta+\ga^2}$ although the parameters $p_i$ in
\r{b4}-\r{b6}
then are infinite. For $\del\ra0$ we  find the limiting values to be
\be
&&R_1=-{2i\ga^2\la_1B\over{\al\beta+\ga^2}}-{i\al
A(\beta\la_1+\al\la_2+\ga\la_3)\over\al\beta+\ga^2},\nn\\
&&R_2={i\ga(\beta\la_1+\al\la_2)(B-A)\over{\al\beta+\ga^2}}-
{i\ga^2A\la_3\over{\al\beta+\ga^2}}
+\nn\\&&+{B\over2(\al\beta+\ga^2)}\left\{(2\al\beta^2+2\beta\ga^2)\la_1+
(\al^2+\al\beta+2\al\ga^2)\la_2+(2\al\beta\ga+2\ga^3)\la_3\right\},\nn\\
&&R_3={i\ga(\beta\la_1+\al\la_2)(B-A)\over{\al\beta+\ga^2}}-
{i\ga^2A\la_3\over{\al\beta+\ga^2}}
-\nn\\&&-{B\over2(\al\beta+\ga^2)}\left\{(2\al\beta^2+2\beta\ga^2)\la_1+
(\al^2+\al\beta+2\al\ga^2)\la_2+
(2\al\beta\ga+2\ga^3)\la_3\right\},\nn\\
&&R_4={2i\ga^2\la_2B\over{\al\beta+\ga^2}}+{i\beta
A(\beta\la_1+\al\la_2+\ga\la_3)\over\al\beta+\ga^2},
\e{b81}
where $A$ and $B$ is given in \r{a06}-\r{a07} appendix A. At
$\del\ra\pm2\sqrt{\al\beta+\ga^2}$ we find
\be
&&R_{1}=-{\la_2i\sin2\sqrt{\al\beta+\ga^2}\over\sqrt{\al\beta+\ga^2}}, \nn\\
&&R_{2}=(\beta\la_1+\al\la_2+\ga\la_3)
{\sin^2\sqrt{\al\beta+\ga^2}\over\al\beta+\ga^2}-
i\la_3{\sin2\sqrt{\al\beta+\ga^2}\over
\sqrt{\al\beta+\ga^2}},\nn\\
&&R_{3}=-(\beta\la_1+\al\la_2+\ga\la_3){\sin^2\sqrt{\al\beta+\ga^2}
\over\al\beta+\ga^2}-
i\la_3{\sin2\sqrt{\al\beta+\ga^2}\over\sqrt{\al\beta+\ga^2}},\nn\\
&&R_{4}={\la_2 i\sin2\sqrt{\al\beta+\ga^2}\over\sqrt{\al\beta+\ga^2}}.
\e{b82}

   That the remaining
basic variables transform exactly as in
\r{a2} follows trivially from the algebra since
the $V_k$ generators only couples to
$x^a$ and $v^a$. The transformation of the basic
generators $K_i$ and $V_k$ follows
readily, (Below, $K'_i$, refers to the $SL(2,{
R})\times U(1)$-transformed part, given by \r{a3}.)
\be
&K''_1=&K'_1+(A-i\gamma B)R_3e^{i\frac{\delta}{2}}V_1
-i\beta B R_4e^{i\frac{\delta}{2}} V_2+\nn\\
&& +((A-i\gamma B)R_4+i \beta B R_3)e^{i\frac{\delta}{2}}V_3,\nn\\
&K''_2=&K'_2+i\alpha BR_1e^{i\frac{\delta}{2}} V_1
-i(A+i\gamma B)R_2e^{i\frac{\delta}{2}} V_2+\nn \\
&& +((A+i\gamma B)R_2+i\alpha B R_2)e^{i\frac{\delta}{2}}V_3,\nn\\
&K''_3=&K'_3+(i\alpha B R_3 -(A-i\gamma B)R_1)e^{i\frac{\delta}{2}}
V_1-\nn\\ &&-(A+i\gamma B)(iR_4-i\beta B R_2)
e^{i\frac{\delta}{2}} V_2+\nn
\\ && +((A+i\gamma B)R_4+i\alpha B R_4-i\beta BR_1-
(A-i\gamma B)R_2)e^{i\frac{\delta}{2}}V_3,\nn\\
&K''_4=&K'_4+(i\alpha B R_3-(A-i\gamma
B)R_1)e^{i\frac{\delta}{2}}V_1-\nn\\ &&-((A+i\gamma B)R_4-i\beta B
R_2)e^{i\frac{\delta}{2}}V_2+\nn\\ &&+((A+i\gamma B)R_4+i\alpha B
R_4-i\beta B R_1- (A-i\gamma B)R_2)e^{i\frac{\delta}{2}}V_3,\nn\\
&V''_1=&({(A-i\gamma B)}^2 V_1-{\beta}^2 B^2 V_2+2i\beta B(A-i\gamma
B)V_3)e^{i\delta},\nn\\
&V''_2=&({(A+i\gamma B)}^2 V_2-2i\alpha B(A-i\gamma B)V_1-i\beta
B(A+i\gamma B)V_2)e^{i\delta},\nn\\
&V''_3=&((A^2+{\gamma}^2B^2-\alpha\beta B^2)V_3+i\alpha
B(A-i\gamma B)V_1-i\beta B(A+i\gamma B)V_2)e^{i\delta}.
\e{b9}
\vspace{5mm}\\
{\bf Derivation of the reduction formulas \r{47}.}\\ \\
In order to compute the action of the regulator factor \r{b1} on
the states $|\phi\hb_1$ and $|\phi\hb_2$ it is convenient to use
other factorizations than \r{b2}-\r{b3}.
Two different factorizations  suitable for actions on
$|\phi\hb_1$ and $|\phi\hb_2$ follow from \r{b61} and \r{a9}-\r{a10}.
They are given
by
\be
&&e^{[Q,\psi]}=e^{ip_3 V_3}e^{ip_1 V_1}e^{ip_2 V_2}
e^{f_1 K_1}e^{(f_2+if_3)K_3}e^{f_4 K_2}e^{\delta K_4}
e^{-ip_2 V_2}e^{-ip_1 V_1}e^{-ip_3V_3},
\e{b11}
\be
&&e^{[Q,\psi]}=e^{ip_3 V_3}e^{ip_1 V_1}e^{ip_2 V_2}
e^{g_1 K_2}e^{(g_2+ig_3)K_3}e^{g_4 K_1}
e^{\delta K_4}e^{-ip_2 V_2}e^{-ip_1 V_1}e^{-ip_3V_3},
\e{b12}
where the parameters $p_i$ are given by \r{b4}-\r{b6}, and
where $f_i$ and $g_i$ are
given in \r{a11} and \r{a12} respectively.
The factorizations \r{b11} and \r{b12} imply
\be
&&e^{[Q,\psi]}|\phi\hb_1=
e^{f_1 K_1}e^{{\mu V_1}}
|\phi\hb_1, \quad e^{[Q,\psi]}|\phi\hb_2=
e^{g_1 K_2}e^{\nu V_2}|\phi\hb_2,
\e{b15}
where
\be
&&\mu\equiv -ie^{\{i\delta-2i(f_2+i
 f_3)\}}p_1+ip_1+f_1p_3+i{f_1}^2 p_2,\nn\\
&&\nu\equiv -ie^{\{i\delta-2i(g_2+i
 g_3)\}}p_2-ip_2+g_1p_3-i{g_1}^2 p_1.
\e{b16}
Note that $\mu$ and $\nu$ are finite at $\del\ra0$ and
$\del\ra\pm\sqrt{\al\beta+\ga^2}$ although $p_1, p_2$ and
$p_3$ are infinite in these limits. Since $\mu$ and $\nu$ may be introduced by
unitary gauge transformations  they do not affect the
inner product properties (see
below and  section 6).
 \vspace{5mm}\\
{\bf Conditions for finite inner-products}\\ \\
The  non-hermitian operator doublets are here given by
\be
D_{(1)}'= e^{[Q,\psi]}D_{(1)}e^{-[Q,\psi]} = e^{f_1
K_1}e^{\mu V_1}D_{(1)}e^{-\mu V_1}e^{-f_1 K_1}\nn\\
=\{x^a-if_1v^a-2i\mu
p^a,\ca^a-if_1{\bar{\pet}}^a,{\bar{\ca}}^a+
if_1\pet,\pi^a+if_1p^a\},
\e{b17}
\be
D_{(2)}'= e^{[Q,\psi]}D_{(2)}e^{-[Q,\psi]} = e^{g_1
K_2}e^{\nu V_2}D_{(2)}e^{-\nu V_2}e^{-g_1 K_2}\nn\\
=\{v^a-ig_1x^a-2i\nu
\pi^a,\pet^a+ig_1{\bar{\ca}}^a,
{\bar{\pet}}^a-ig_1\ca^a,p^a+ig_1\pi^a\},
\e{b18}
where
\be
D_{(1)}=\{x^a,\ca^a,{\bar{\ca}}_aa,\pi_a\}, \quad
D_{(2)}=\{v^a,\pet^a,{\bar{\pet}}_a,p_a\}.
\e{b19}
We find then
\be
Det([D'_{(1)},{D'_{(1)}}^{\dagger}])=
\left |
\left(
\begin{array}{cccc}
-4 Re(\mu) & 0 & 0 & 2Re(f_1)\\
0 & 0 & -2iRe(f_1) & 0\\
0 & 2iRe(f_1) & 0 & 0\\
2 Re(f_1) & 0 & 0 & 0\\
\end{array}
\right)
\otimes \bf{1_{n\times n}}
\right|\nn
\ee
\be
={(16 Re(f_1))}^{4n},
\e{b20}
and similarly
\be
&&Det([D'_{(2)},{D'_{(2)}}^{\dagger}])={(16 Re(g_1))}^{4n}.
\e{b21}
The basic condition \r{7} for inner product spaces require then
\be
&&Re f_1\neq0\;\Leftrightarrow \;\al\neq0,\;F\neq0,\nn\\
&&Re g_1\neq0\;\Leftrightarrow \;\beta\neq0,\;F\neq0,
\e{b22}
where $F$ is defined in \r{603}. Thus
\be
&&F\neq 0\;\Leftrightarrow\;\al\beta+\ga^2\neq
\frac{n^2\pi^2}{4}\;\;\mbox{for any
integer  } n\neq0.
\e{b23}
These conditions are identical to those in section 2.

\end{document}